\documentclass[11pt,a4paper]{article}
\pdfoutput=1
\usepackage[latin1]{inputenc}
\usepackage{amsfonts,amsbsy,bm,euscript,mathrsfs}
\usepackage{amssymb,stmaryrd,faktor,slashed}
\usepackage{color}
\usepackage[tbtags]{amsmath}
\usepackage[bookmarks=true,colorlinks=true,linkcolor=black,citecolor=black,urlcolor=black,bookmarksnumbered]{hyperref}
\usepackage[nodayofweek]{date time}

\usepackage[parsep]{collref}

\usepackage[nosort]{cite}

 \usepackage[a4paper,text={170mm,257mm},centering]{geometry}

 \usepackage{authblk, physics, bbold, multirow, comment}
\usepackage{graphicx}
\usepackage[verbose]{wrapfig}
\usepackage{caption}

\numberwithin{equation}{section}

\makeatletter
\renewcommand\section{\@startsection {section}{1}{\z@}%
	{-3.5ex \@plus -1ex \@minus -.2ex}%
	{2.3ex \@plus.2ex}%
	{\normalfont\large\bfseries}}
\renewcommand\subsection{\@startsection{subsection}{2}{\z@}%
	{-3.25ex\@plus -1ex \@minus -.2ex}%
	{1.5ex \@plus .2ex}%
	{\normalfont\normalsize\bfseries}}
\makeatother

\expandafter\def\expandafter\bfseries\expandafter{\bfseries\ifmmode\else\boldmath\fi}
\expandafter\def\expandafter\mdseries\expandafter{\mdseries\ifmmode\else\unboldmath\fi}
\expandafter\def\expandafter\normalfont\expandafter{\normalfont\ifmmode\else\unboldmath\fi}

\providecommand{\href}[2]{#2}

\newcommand{\mathsym}[1]{{}}
\def\id{\protect{{1 \kern-.28em{\rm l}}}}
\def\be{\begin{equation}}
\def\ee{\end{equation}}

\def\tr{{\rm tr\,}}
\def\ha{\tfrac{1}{2}}
\def\td{\tilde}
\def\ci{\cite}
\def\N{{\mathcal N}}

\def\a{\alpha}

\def\g{\gamma}

\def\l {\lambda}

\def\foot{\footnote}
\newcommand{\rf}[1]{(\ref{#1})}

\def\no{\nonumber}

\def\la{\label}
\def\l{\lambda}

\def\adss{$AdS_5 \times S^5$\ }

\def\r{\rho}

\def\varpi{{\rm w}}

\def\R{{\rm R}}

\def\del{\partial}

\def\edd{\end{document}}

\def\iffa{\iffalse}

\def\C {{\rm C}}

\def\dd {{\rm d}}

\def\vp{\varphi}

\def \ov {\over}

\def \dt {\dt}

\DeclareMathOperator\vol{vol}

\def \ZZ {\mathbb Z}

\def \iffa {\iffalse}

\def \adss {$AdS_5 \times S^5$ } \def \N {{\cal N}} \def \ZZ {\mathbb Z}
\def \a {\alpha}
\def\ov{\over}
\def \ci {\cite}

\def \foot {\footnote}

\def\la{\label}
\def\foot{\footnote}
\def \OO {{\cal O}} \def \no {\nonumber}
\def \ed {\end{document}}
\def \adss {$AdS_5 \times S^5$\ }
\def \adsz {$AdS_5 \times S^5/\ZZ_2$\ }
\def\ba#1\ea{\begin{align}#1\end{align}}	
\def \ssp {\nabla^2_{_{AdS_5}}}
\def \sstar {*}

\def \M {{\cal M}}
\def \A {p} \def \B {q}
\def \VV {{\rm V}}
\def \bbeta {\varphi} \def \gf {{\tilde f_3}} 
\def \Vv {{\cal V}} \def \NN {{\rm n}} \def \Ca {{C}}
\def \R {{\cal R}} \def \C {{\cal C}} 

\def \nun {\nu }

\begin{document}

\begin{flushright}\small{Imperial-TP-AT-2023-{07}}
\end{flushright}
\vspace{0.1cm}
\begin{center}
			{\Large\bf On AdS/CFT duality in the twisted sector of \\ 
			\vspace{0.1cm}
			 string theory on $AdS_5 \times S^5/\ZZ_2$ orbifold background
		 
			\vspace{0.1cm}
		}
		\vspace{1.0cm}		
		{Torben Skrzypek$^{a,}$\footnote{t.skrzypek20@imperial.ac.uk} and
			Arkady A. Tseytlin$^{a,}$\footnote{Also at the Institute for Theoretical and Mathematical Physics (ITMP) of Moscow U. and
				Lebedev Institute.
				
				\ \ tseytlin@imperial.ac.uk}
		}	
		
		\vspace{0.3cm}		
		{\em
			$^{a}$Theoretical Physics Group, Blackett Laboratory
			\vspace{0.1cm} \\ Imperial College, London SW7 2AZ, U.K.
		}		
\end{center}	
\vspace{0.5cm}
	
\begin{abstract}
We consider type IIB string theory on an 
$AdS_5 \times S^5/\mathbb{Z}_2$ orbifold background, 
which should be dual to 4d $\mathcal{N}=2$ superconformal $SU(N)\cross SU(N)$ 
 gauge theory with two bi-fundamental hypermultiplets.
 The correlator of two chiral BPS operators from the twisted sector of this quiver CFT exhibits non-trivial dependence on the 't Hooft coupling $\lambda$ already in the planar limit. 
 This dependence was recently determined using localisation 
 and the expansion 
 at large $\lambda$ contains a subleading 
 contribution proportional to $\zeta(3) \lambda^{-3/2}$. 
 We address the question of how to reproduce this correction on the string theory side 
 by starting with the $\zeta(3) \alpha'^3$ term in the type IIB string effective action. 
 We find a solution of type IIB supergravity 
 which represents a resolution of the $AdS_5 \times S^5/\mathbb{Z}_2$ orbifold 
  singularity
  and demonstrate that 
 the relevant light twisted sector states may be identified as additional supergravity 2-form modes 
 ``wrapping" a finite 2-cycle in the resolution space. 
 Reproducing the structure of the 
 gauge theory result becomes more transparent in the large $R$-charge or BMN-like limit 
 in which the resolved background takes a pp-wave 
 form with the transverse space being a product of $\mathbb R^4$ and the Eguchi-Hanson space. 
 \end{abstract}

\newpage
\tableofcontents
\newpage

\setcounter{footnote}{0}

\section{Introduction}\label{intro}

One of the simplest generalisations of the duality between $\N=4$ SYM theory and type IIB superstring theory on \adss background 
is based on taking its orbifold \cite{Kachru:1998ys} (see also, e.g., 
\cite{Lawrence:1998ja,Bershadsky:1998mb,Bershadsky:1998cb,Gukov:1998kk,Klebanov:1998hh,Klebanov:1999rd,Gadde:2009dj}). 
In particular, in the case of a supersymmetric 
$\ZZ_2$-orbifold the duality is between the 
$\N=2$ superconformal $SU(N) \times SU(N)$ quiver gauge theory (containing two vector multiplets and
 two bi-fundamental hypermultiplets)
and string theory on $AdS_5 \times S^5/\ZZ_2$. 

A way to check this duality is to compute ``observables" on the gauge theory side for large $N$ and any 't Hooft coupling $\lambda$, expand in large $\l$ and then compare the result to the large-tension expansion of their counterparts 
 on the string theory side. 
 As the $\N=2$ superconformal quiver 
 arises as a $\mathbb{Z}_2$-orbifold of $SU(2N)$ $\mathcal{N}=4$ SYM 
 \cite{Lawrence:1998ja}, its ``untwisted" sector observables, computed at the leading (planar) 
 order in large $N$, are the same as in the SYM theory\foot{In general, to construct an orbifold 
 one starts with a discrete subgroup $\Gamma$ of the $PSU(2,2|4)$-symmetry. 
 $\Gamma$-invariant states form the untwisted sector. 
 In string theory, additional twisted sector states 
 arise from strings that close only up to a $\Gamma$-transformation. These states correspond in the dual 
 gauge theory to operators with an insertion of a $\Gamma$-symmetry operator.}
 with non-trivial corrections appearing at order $1/N^2$.

Recently, such leading $1/N^2$-corrections were studied for some  
 simple ``untwisted"
  observables -- the BPS circular Wilson loop and the free energy on the 4-sphere 
 (see \cite{Beccaria:2021ksw, Beccaria:2022ypy,Beccaria:2023kbl} and references therein). Localisation \cite{Pestun:2007rz} allows one 
 to compute their strong 't Hooft coupling expansion order by order in $1/N^2$. 
 The comparison to string theory at order $1/N^2$ then requires knowledge of string loop corrections 
 in $AdS_5 \times S^5/\ZZ_2$
 which, unfortunately, is rather limited (see a discussion in \cite{Beccaria:2021ksw,Beccaria:2023kbl}).

At the same time, observables from the twisted sector may receive corrections already at 
the leading order in large $N$.
These should be captured by the tree-level string theory 
 and thus may be easier to analyse. 
 In particular, the 
 2- and 3-point correlators of special twisted sector single-trace BPS operators 
 with protected dimensions\foot{At the leading order in large $N$ the 
spectrum of anomalous dimensions of non-BPS states may, in principle, be studied using integrability techniques
(see \cite{Skrzypek:2022cgg} and references therein).} 
 may be 
 computed at large $N$ using localisation techniques 
 (see, e.g., 
\ci{Pini:2017ouj,Billo:2017glv,Beccaria:2020hgy,
Galvagno:2020cgq,Beccaria:2021hvt,
Fiol:2021icm,Billo:2021rdb,Billo:2022gmq,Billo:2022fnb,Billo:2022lrv,Beccaria:2023kbl}).
The leading order large-$\l$ terms in these correlators 
were matched \ci{Billo:2022gmq,Billo:2022fnb,Billo:2022lrv}
with the 6d 
low-energy effective action for the corresponding 
twisted sector string 
modes constructed in \cite{Gukov:1998kk}.

Our aim below is to attempt 
to extend this matching to the first sub-leading $\order{\lambda^{-3/2}}$-term in
 the 2-point twisted-state correlator 
\cite{Billo:2022lrv}. 
We expect this term to be captured by the $\alpha'^3$-correction to the string effective action.

\subsection{Gauge theory results}

If $\phi_0$ and $\phi_1$ denote the adjoint scalars in the two $\N=2$ $SU(N)$ 
vector multiplets, 
the simplest chiral BPS operators of dimension $\Delta=k$ belonging to untwisted and 
twisted sectors ,respectively, are\foot{We use a different normalisation than in \cite{Billo:2022lrv}.} 
\begin{equation}\label{ops}
U_k(x)=\frac{1}{\sqrt{2k }}\Big(\frac{2}{N}\Big)^{\frac{k}{2}}\Big(\tr \phi_0^k+\tr\phi_1^k\Big)\,,\qquad \qquad T_k(x)=\frac{1}{\sqrt{2k }}\Big(\frac{2}{N}\Big)^{\frac{k}{2}}\Big(\tr \phi_0^k-\tr\phi_1^k\Big)\,. 
\end{equation}
 As their conformal dimension is protected, the corresponding 
 2- and 3-point correlators are \cite{Billo:2022lrv}
\ba \langle \mathcal{O}_k(x_1)\bar{ \mathcal{O}}_k(x_2)\rangle&=\frac{{\rm G}_{\mathcal{O}_k}}{\abs{x_1-x_2}^{2k}}\,, \la{1.1}\\
\langle \mathcal{O}_k(x_1)\mathcal{O}_l(x_2)\bar{\mathcal{O}}_{k+l}(x_3)\rangle&=\frac{{\rm G}_{\mathcal{O}_k,\mathcal{O}_l,\bar{ \mathcal{O}}_{k+l}}}{\abs{x_1-x_3}^{2k}\abs{x_2-x_3}^{2l}}\,,\la{2}
\ea 
where the operator $\mathcal{O}_k$ is either $U_k$ or $T_k$ in \rf{ops} and the constants $\rm G$ may 
 depend on $N$ and the 't Hooft coupling $\lambda$. 

\begin{sloppypar} 
We shall focus on the leading order in large $N$. Then the correlators of untwisted BPS operators are the same
 as in $\mathcal{N}=4$ SYM theory, i.e. they 
are protected by supersymmetry and are given by 
\begin{equation}\la{5}
{\rm G}_{U_k}=1\,,\qquad \qquad \ \ {\rm G}_{U_k,U_l,\bar{U}_{k+l}}=\frac{1}{N} \sqrt{\frac{kl(k+l)}{2}}\,.
\end{equation} 
On the other hand, the correlators involving twisted sector operators are non-trivial functions of $\l$ that can be found using localisation. 
 Expanding in large $\l$ one gets, in particular (see \cite{Galvagno:2020cgq,Beccaria:2021hvt,Billo:2021rdb,Billo:2022gmq} and \cite{Billo:2022fnb,Beccaria:2022ypy,Billo:2022lrv,Beccaria:2023kbl})\footnote{\la{f4}Eqs.\eqref{3} and \eqref{4} are 
 taken from 
 \cite{Billo:2022lrv}. The leading large-$\l$ coefficients 
 were found 
 earlier in \cite{Billo:2021rdb,Billo:2022fnb}. The resummation in terms of $\l'$ was introduced in \ci{Beccaria:2022ypy}. 
 Ref. \cite{Billo:2022lrv} found the following 
 relation between the 2- and 3-point coefficient functions:
	\begin{equation*}
	{\rm G}_{T_k,T_k,\bar{U}_{2k}}=\frac{k}{\sqrt{2}}\Big(\frac{N}{2}\Big)^{k-1}(k+\lambda\partial_\lambda){\rm G}_{T_k}\,.
	\end{equation*}
 }
 \begin{align}
&{\rm G}_{T_k} =\frac{4\pi^2}{\lambda'}k(k-1)\Big(\frac{\lambda'}{\lambda}\Big)^{k}\Big[1
+\tfrac{1}{2}(2k-1) (2k-2)(2k-3) \frac{\zeta(3)}{\lambda'^{\frac{3}{2}}} \no \\ 
\label{3} &\qquad\qquad\qquad\qquad \qquad -\tfrac{9}{16} (k-1) (2k-3)(2k-5)(4 k^2 - 1) \frac{\zeta(5)}{\lambda'^{\frac{5}{2}}} \\ &\qquad\qquad\qquad\qquad \qquad 
+\tfrac{1}{4} (k-1)(2k-1)(2k -3)(2k-5)(4k^2 -20k-3) \frac{\zeta(3)^2}{\lambda'^{3} } 
+\OO\big({\lambda'^{-{7/2}}}\big)\Big]\,,\no \\
&{\rm G}_{T_k,T_k,\bar{U}_{2k}}=\frac{k^{\frac{3}{2}}}{N}\cdot\frac{4\pi^2}{\lambda'}(k-1)^2\Big(\frac{\lambda'}{\lambda}\Big)^{k-\frac{1}{2}}\Big[1+\tfrac{1}{2} (2k-1)(2k-3)(2k-5)\frac{\zeta(3)}{\lambda'^{\frac{3}{2}}}+\OO\big({\lambda'^{-{5}/{2}}}\big)\Big]\,,\la{4}\\ &\qquad\qquad\qquad\qquad \qquad \sqrt{\lambda'}\equiv \sqrt{\lambda}-4\log 2 \ . \la{44}
\end{align}
 In general, these expressions depend on operator normalisations. 
 Considering normalisation-independent ratios like
 $\tfrac{{\rm G}_{T_k,T_k,\bar{U}_{2k}}}{\sqrt{{\rm G}_{T_k}{\rm G}_{T_k}{\rm G}_{U_k}}}$,
 refs. \cite{Billo:2022gmq,Billo:2022fnb} 
 successfully matched their leading large-$\lambda$ behaviour to the predictions
 from the low-energy effective action \cite{Gukov:1998kk} 
 for the corresponding 
 twisted sector string modes.
\end{sloppypar}

We shall attempt to understand the string origin of the subleading $\zeta(3)$-term in \rf{3}. 
Identifying $\sqrt{ \l'}\ov 2 \pi $ with the effective string tension or $L^2\ov 2\pi \a'$ 
(where $L$ is the radius of both $AdS_5$ and $S^5$), 
the natural expectation is that this term should 
be reproduced
 by the first non-trivial $\a'^3$-correction      
  in the string 
 effective action for the corresponding 
 twisted sector modes in $AdS_5 \cross S^5/{\mathbb Z}_2$,
  by analogy with 
  the familiar $\a'^3 \zeta(3)\mathcal{R}^4+...$ term for the standard massless string modes (cf. \ci{Banks:1998nr,Gubser:1998nz}). 
This   will require understanding 
how to construct   a  generalisation of the leading-order 
 effective action for the twisted sector modes suggested in \cite{Gukov:1998kk}. 

\subsection{6d effective action for twisted sector modes}

To recall, 
the $\Gamma=\ZZ_2$ orbifold on the string theory side acts on embedding coordinates $(z_1,z_2,z_3)$ of $S^5\subset\mathbb{C}^3$ as
\begin{equation}\la{1.6}
\Gamma:\quad(z_1,z_2,z_3)\to(-z_1,-z_2,z_3)\,, 
\end{equation}
which breaks half of the maximal supersymmetry. The great circle of $S^5$ parametrised by
$z_1=z_2=0, \ z_3= L\, e^{i\chi}$
is fixed under the action of $\Gamma$. 
The twisted sector strings, which close up to a $\Gamma$-transformation, extend around the orbifolded angles, so the lowest energy twisted states localise on the fixed circle, i.e. 
 on the six-dimensional $AdS_5\cross S^1$ subspace of the 10d target space. They should then be described by an effective 6d action.

For the similar orbifold $\mathbb{R}^{1,5}\cross \mathbb{C}^2/\Gamma$ of flat-space string theory 
 the spectrum of twisted sector 
modes organises into tensor representations of 6d $\mathcal{N}=(2,0)$ supersymmetry \cite{Douglas:1996sw}
(see appendix \ref{appendix A}). The low-energy effective action for the light twisted modes 
 could be reconstructed from correlators of the corresponding vertex operators (cf. \ci{Berenstein:2000hy}). 

Ref. \cite{Douglas:1996sw} provided an alternative interpretation of the twisted sector modes 
 in terms of a resolution (or ``blow-up") of the orbifold singularity. 
One may cut out a ball of size $a$ around the singularity and glue in a smooth manifold, such that the total space
 $\mathcal{M}^4$ 
is asymptotically locally Euclidean with global $\mathbb{C}^2/\Gamma$ structure. 
For $\Gamma=\mathbb{Z}_2$ the smooth manifold is the Eguchi-Hanson space \cite{Eguchi:1978gw}. 
This resolution features three moduli and a non-trivial 2-cycle over which one may integrate the massless 
2- and 4-form fields of type IIB supergravity generating extra light modes. 
 In the limit $a\to 0$, the resolved space $\mathcal{M}^4$ approaches the orbifold $\mathbb{C}^2/\mathbb{Z}_2$ with 
 the moduli and extra modes (now localised at the singularity) to be taken into account. 
 It turns out that this procedure 
 reproduces the lightest states in the twisted sector spectrum as found directly 
 from string theory.
 This suggests that one can access the light twisted sector modes 
 using the 10d supergravity action expanded near the curved 
 background representing a resolution of the orbifold. 

In \cite{Gukov:1998kk} it was suggested that this logic may apply also 
 to the curved-space orbifold $S^5/\mathbb{Z}_2$. 
 Close to the fixed circle, one may approximate $S^5/\mathbb{Z}_2$ by $ \mathbb{C}^2/\mathbb{Z}_2\cross S^1$
 and thus expect to get the same effective action for light twisted sector modes in terms of 6d tensor 
 multiplets as in the flat-space case, corrected by contributions of the 
 curvature and the $F_5$-flux of the \adsz background. 
 One may then expand in Fourier modes on $S^1$, generating 
 towers of fields in $AdS_5$ with masses
 labelled by the mode number (``KK level'') $k$. 
 These can then be put into correspondence with the dual BPS operators in the twisted 
 sector of the gauge theory 
 and turn out to have the required spectrum of conformal dimensions
 \cite{Gukov:1998kk}.

 In particular, the twisted sector operator $T_k$ in \eqref{ops} is expected 
 to be dual to a 5d mode representing a combination of $B_2$ and $C_2$ fields integrated over 
 the 2-cycle the orbifold resolution. 
 The relevant terms in the 10d type IIB supergravity action are (ignoring dependence on the dilaton and RR scalar) 
 \begin{equation}\la{0}
S_{10}=-\frac{1}{2\kappa^2}\int\dd ^{10} X\,\sqrt{-g} \,\bigg(\frac{1}{2\cdot3!}H_3^2+\frac{1}{2\cdot 3!}F_3^2+\frac{1}{4\cdot 5!}\tilde F_5^2\bigg)-\frac{1}{4\kappa^2} \int \, B_2\wedge F_3\wedge F_5\,,
\end{equation}
\begin{equation}
 H_3= \dd B_2\ , \ \qquad F_3 = \dd C_2 \ , \qquad \tilde F_5=F_5-\frac{1}{2}C_2\wedge H_3+\frac{1}{2} B_2\wedge F_3\,.\la{00}
\end{equation}
Let us set 
\begin{equation}
B_2=\beta(x,\chi)\, \Theta\,,\qquad \qquad C_2=\gamma(x,\chi)\, \Theta\, , \la{17}
\end{equation}
where $\Theta$ is the anti-self-dual 
2-form on the resolution of 
$\mathbb{C}^2/\mathbb{Z}_2$ \cite{Eguchi:1978gw} 
with a normalised integral over the resolution 2-cycle, $\chi$ is the fixed $S^1$ coordinate and $x^i$ are $AdS_5$ coordinates. 
 Using that $F_5= 4 \,(\vol_{AdS_5}+ \vol_{S^5/\mathbb{Z}_2})$
we then arrive at the following effective 6d action for the fields $\beta$ and $\gamma$ 
\ci{Gukov:1998kk,Klebanov:1999rd} 
\begin{equation}
S_6\sim\ha\int_{AdS_5\cross S^1}\dd^5 x\,\dd\chi\,\sqrt{-g_6}\,\Big[(\partial_i\beta)^2+(\partial_\chi\beta)^2+(\partial_i \gamma)^2+(\partial_\chi\gamma)^2-8\, \beta\,\partial_\chi\gamma \Big]\,. \la{18} 
\end{equation}

Expanding the fields 
$\beta$ and $\gamma$ in Fourier modes in $\chi$ (i.e. $\beta = \sum_k e^{i k \chi} \beta_k(x)$, $\g = \sum_k e^{i k \chi} \g_k(x)$)
we get the following kinetic operator matrix for the $\beta_k(x)$ and $\gamma_k(x) $ fields on $AdS_5$
\begin{equation}
\begin{pmatrix}
{\ssp}-k^2 & -4i k \\
4ik & {\ssp}-k^2
\end{pmatrix} \ . \la{19}
\end{equation}
 Diagonalising it gives the following masses 
\begin{equation}
m^2_\pm =\Delta_\pm (\Delta_\pm -4)=k(k\pm 4)\,. \la{20} 
\end{equation}
The dual twisted-sector operators are then expected to be $T_k$ in \eqref{ops} with 
dimension $\Delta_- =k$ and 
the operator with dimension 
$\Delta_+ = k+4$ represented by \cite{Gukov:1998kk} 
\begin{equation}
\mathcal{O}_F=\tr\big[\phi_0^k\, (F_0^2+iF_0\tilde F_0)\big]-\tr\big[\phi_1^k\, (F_1^2+iF_1\tilde F_1)\big]\,, \la{21}
\end{equation}
where $F_0$ and $F_1$ are the gauge fields from the two $\N=2$ $SU(N)$ gauge multiplets. 

This identification relies heavily on supersymmetry 
and is supported by the 
successful matching of the leading order term in \eqref{3} demonstrated in \cite{Billo:2022fnb}.
To extend this matching to subleading order we need to put the 
above derivation of the effective action 
 \rf{18} on a firmer footing and then 
 find $\a'^3$-corrections to it using as an input the known structure of $\a'^3$-terms 
 in the type IIB string effective action. 
 
 \subsection{Structure of the paper}

We start 
 in section \ref{sec 2} with presenting a solution of type IIB supergravity 
(depending 
 on an extra parameter $a$) that
  represents a resolution of the $S^5/\mathbb{Z}_2$ orbifold singularity and generates a non-trivial 2-cycle. For generic values of $a$ this spacetime is still irregular as we will discuss below, with the exception of special values of $a$ where it becomes either a nontrivial fibration of $S^3$ over $S^2$ \cite{Hashimoto:2004kc,Lu:2004ya} or a true product space $S^3\cross S^2$.
 
In section \ref{sec 3}  we analyse 
solutions for the $B_2$ and $C_2$ fields in this background 
 and identify the 10d 
 analogues of the twisted sector states localised on the $AdS_5 \times S^1$ subspace, 
thus supporting the logic behind the derivation of the action \rf{18} in
 \cite{Gukov:1998kk}. 
 
 In section \ref{sec 4} we generalise the discussion to the case 
 when the starting point is not the supergravity action 
 \rf{0} but the type IIB string effective action including $\a'^3$-corrections. 
 We describe a strategy for reproducing 
 the subleading $\l^{-3/2}$-term in 
 the gauge theory result \eqref{3} for the two-point correlator. 
 
 In section \ref{sec5} we focus specifically on reproducing
 the $k^3$ part of the $ \l^{-3/2}$-term in \rf{3} which dominates when the $R$-charge $k$ is large. 
 We note that the function \rf{3} admits a regular BMN-like limit, i.e. 
 for large $k$ and $\l$ 
 with $\nu = {k\ov \sqrt \l}$ being fixed. 
 This suggests to focus on the pp-wave limit of the resolved orbifold background
 which turns out to have the Eguchi-Hanson space
  as part of its ``transverse''  space.  
We suggest a candidate structure that should be part of the $\a'^3\zeta(3)(R^4 + ...)$ 
superinvariant and that 
may be responsible for reproducing the $\zeta(3) k^3 \l^{-3/2}$-term. 
 
 Some concluding remarks are made in section \ref{sec6}. 
 In appendix \ref{appendix A} we review the spectrum of superstring theory
 on a  flat-space orbifold.
 In appendix
 \ref{appendix B} we summarise some information about 
 the structure of the leading $\alpha'^3$-corrections to the
  tree-level 
  type IIB supergravity action. 
 Details of calculations in section \ref{sec 3} are presented in appendix \ref{appendix C}. 
 Appendix \ref{appendix D} contains expressions for the 
 Weyl tensor 
 of the resolved orbifold background. 

\section{Resolution of the $S^5/\mathbb{Z}_2$ orbifold}\label{sec 2}

In this section we find a particular resolution of the $AdS_5\times S^5/\mathbb{Z}_2$ orbifold
as a solution of type IIB supergravity depending on an extra ``resolution parameter" $a$. 
In section \ref{sec 3} we expand near this background and 
 identify the lightest modes corresponding to the twisted sector states.
 This suggests how to construct their low-energy effective action in the framework of type IIB supergravity. 

 To motivate the Ansatz for the resolved background 
 we first review the resolution of the flat $\mathbb{C}^2/\mathbb{Z}_2$
 orbifold represented by the Eguchi-Hanson (EH) space \cite{Eguchi:1978gw}. 

\subsection{Eguchi-Hanson space as resolution of $\mathbb{C}^2/\mathbb{Z}_2$ }\label{sec 2.1}

The procedure of blowing up singularities
 usually involves glueing a projective $\mathbb{C}{\rm P}^n$ space to the singularity and identifying appropriate subspaces. In the
 $\mathbb{C}^2/\mathbb{Z}_2$ case 
the orbifolding acts on the two complex coordinates as $(z_1,z_2)\to(-z_1,-z_2)$, resulting in a singularity at $(0,0)$. 
Let us choose a parametrisation 
\begin{equation}\label{Euler} 
z_1=r \cos\tfrac{\theta}{2}\ e^{\frac{i}{2}(\psi+\phi)},\qquad z_2=r \sin\tfrac{\theta}{2}\ e^{\frac{i}{2}(\psi-\phi)}\,.
\end{equation} 
Then 
\begin{equation}
\dd s^2 = | dz_1|^2 + |dz_2|^2 = \dd r^2+ r^2(\sigma_x^2+\sigma_y^2+\sigma_z^2)\,,\label{sigma} 
\end{equation}
where we introduced the $SU(2)$ Cartan forms
\begin{equation}\begin{split}
\sigma_x=\tfrac{1}{2}(\sin \psi\,\dd \theta - \sin \theta \cos \psi\, \dd \phi)\,,\,\,
\sigma_y=\tfrac{1}{2}(-\cos \psi\, \dd \theta - \sin \theta \sin \psi \,\dd \phi)\,,\,\,
\sigma_z=\tfrac{1}{2}( \dd \psi + \cos \theta\,\dd \phi)\,,\no 
\end{split}
\end{equation}
\begin{equation}
\qquad
\dd \sigma_x=2 \,\sigma_y\wedge\sigma_z\,,\qquad \dd \sigma_y=2\, \sigma_z\wedge\sigma_x\,,\qquad \dd \sigma_z=2\, \sigma_x\wedge\sigma_y\,.\la{23}
\end{equation}
Here $\sigma_x^2+\sigma_y^2+\sigma_z^2$ represents the metric of $S^3$, parametrised as Hopf fibration
over $S^2$ with $\dd s^2_{S^2}=4(\sigma_x^2+\sigma_y^2)=\dd \theta^2 +\sin^2 \theta\,\dd\phi^2$ and $ \theta\in [0, \pi], \phi \in [0, 2 \pi]$. For 
$\psi \in [0,4\pi]$ it would cover 
 the full $S^3$, but for $\psi \in [0,2\pi]$ it 
 only covers $S^3/\mathbb{Z}_2$, so that 
 \rf{sigma} represents the metric of $\mathbb{C}^2/\mathbb{Z}_2$ with a singularity at $r=0$.

The resolution of the singularity is achieved by replacing \rf{sigma} with the EH metric
containing a function $V_0(r)$, which breaks the $SO(4)$ symmetry to $SO(3)$ 
\begin{equation}\label{244}
\dd s^2_{\text{EH}}= V_0(r)^{-1} \dd r^2 +r^2 \Big[\sigma_x^2+\sigma_y^2+ V_0(r) \,\sigma_z^2\Big]\ , \ \ \ \qquad V_0(r)=1-\frac{a^4}{r^4}\,.
\end{equation}
This metric is Ricci-flat and its curvature form is self-dual.\foot{EH space is also a hyperk\"ahler manifold, i.e. 
 admits three complex structures that form an $SU(2)$ triplet.
In the 10d supergravity context
this guarantees preservation of 
16 supercharges and that a dimensional reduction 
 on this space leads to 6d $\mathcal{N}=(2,0)$ supergravity.} 
 Here we restrict $r$ to the interval 
 $r\in[a,\infty)$. 
 For $r\to\infty$ we recover the $\mathbb{C}^2/\mathbb{Z}_2$ space asymptotically 
 (i.e. EH is an ALE space). 
 For $r\to a$ this space is regular as one can see 
 by changing the coordinate $r\to u$ as 
\begin{equation}
u^2=r^2V_0(r)\,.\la{25}
\end{equation} 
Expanding \rf{244} around the apparent singularity at $r=a$ or $u=0$ 
yields
\begin{equation}\la{26}
\dd s^2_{\text{EH}}\Big|_{u\to 0} \to \frac{1}{4}\Big[ \dd u^2+{a^2}\,\dd s^2_{S^2}+{u^2}(\dd \psi +\cos \theta\,\dd\phi)^2\Big]\,.
\end{equation}
For fixed $\theta$ and $\phi$ and with $\psi\in[0,2\pi]$ the point at $u=0$ is just a coordinate singularity:
 the local geometry near $u=0$ 
is that of an $\mathbb{R}^2$-bundle over $S^2$. 
The orbifold singularity of $\mathbb{C}^2/\mathbb{Z}_2$ is recovered in the limit $a\to 0$.

\subsection{$S^5/\mathbb{Z}_2$ orbifold and its resolution $\M^5$ }\label{sec 2.2}

\begin{figure}[h]
	\centering
	\begin{minipage}{0.49\textwidth}
		\includegraphics[width=\textwidth]{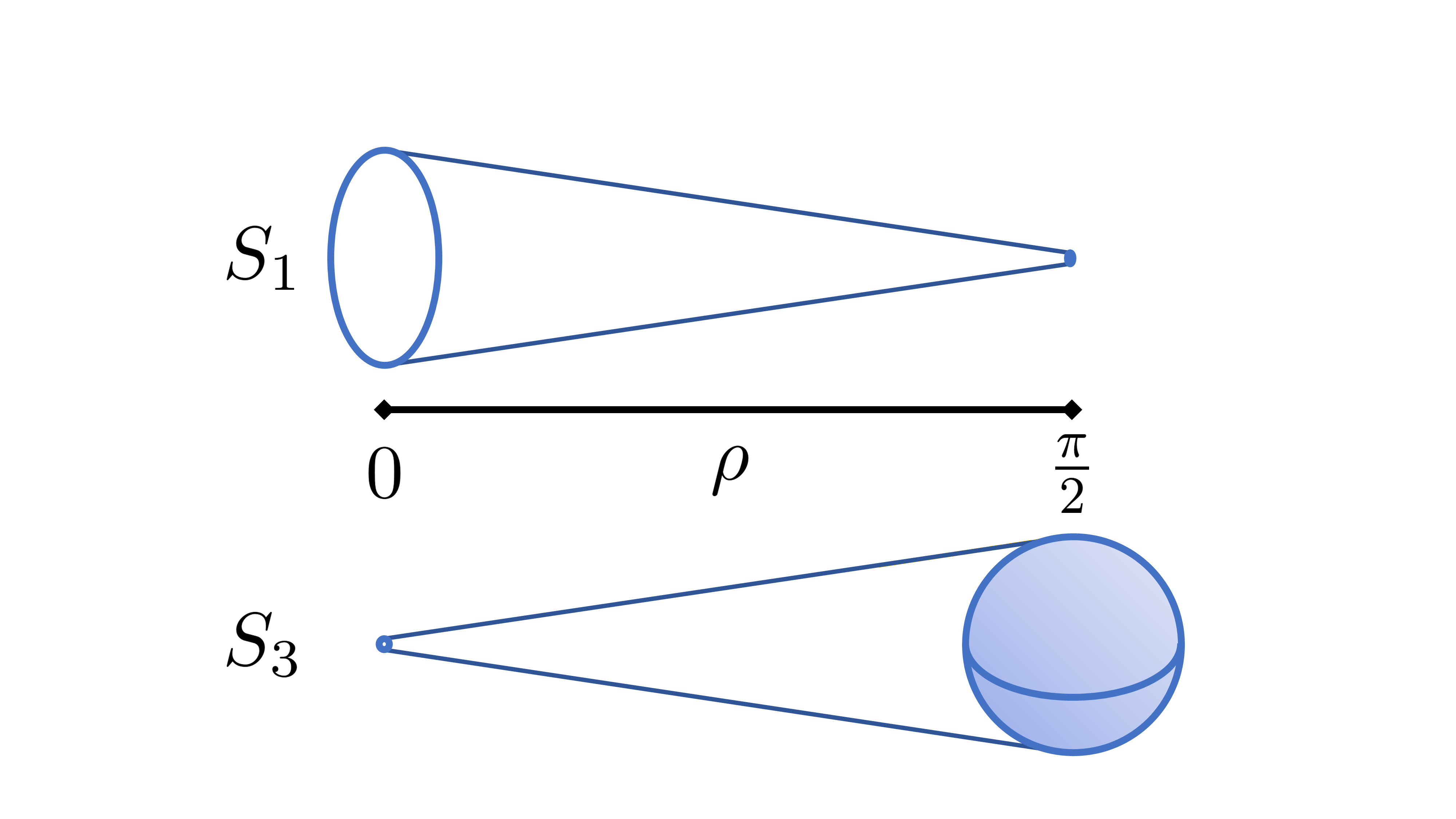}
		\caption*{\textbf{a}}
	\end{minipage}
	\begin{minipage}{0.49\textwidth}
		\includegraphics[width=\textwidth]{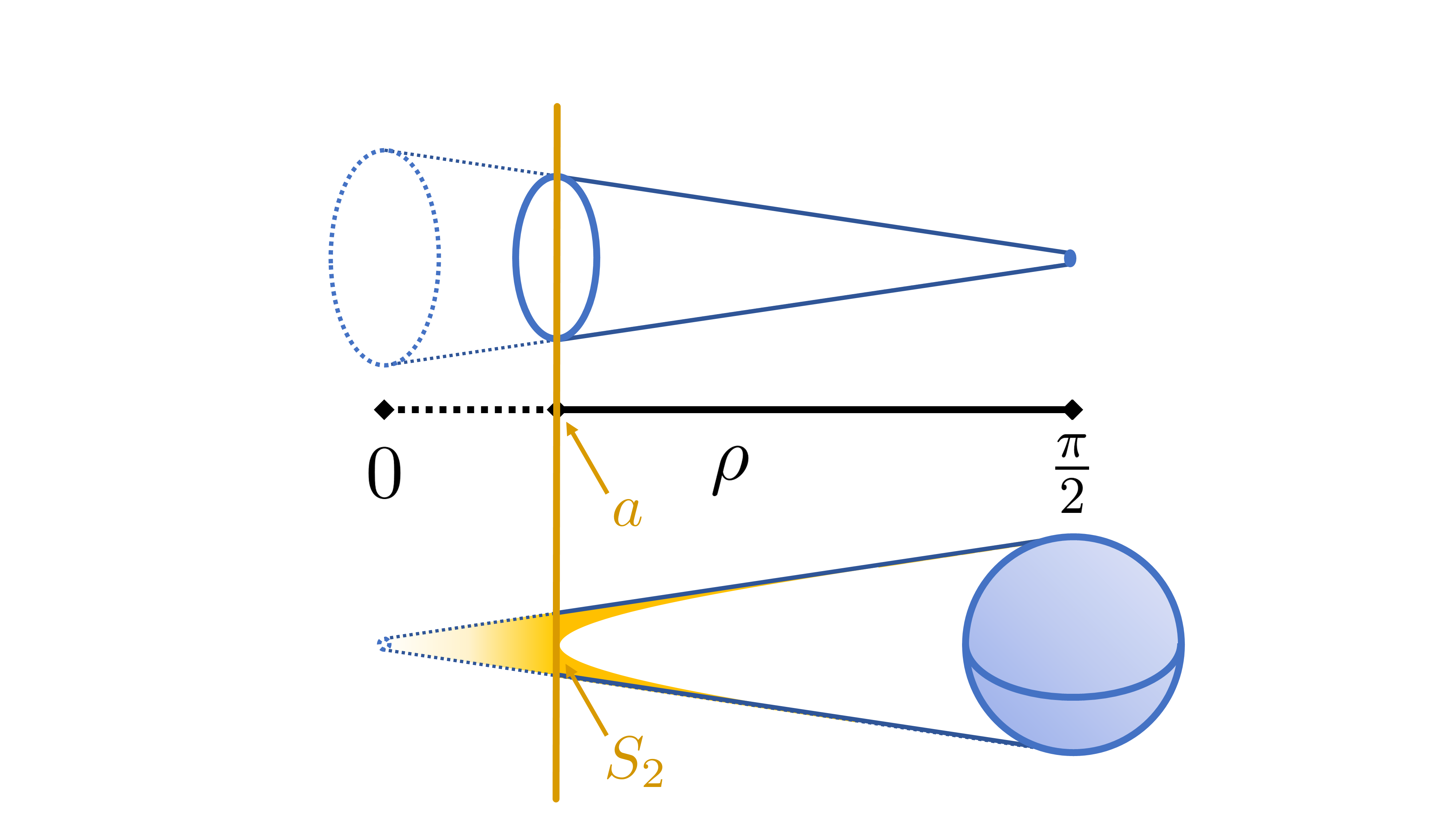}
		\caption*{\textbf{b}}
	\end{minipage}
	\captionsetup{format=hang}
	\caption{\textbf{(a)} Representation of $S^5$ as a fibration of $S^1\cross S^3$ over an interval. \\
			\textbf{(b)} Schematic representation of the resolution $\mathcal{M}^5$ \eqref{xxx}
			of the $S^5/\ZZ_2$ orbifold.}\label{figure1}
\end{figure}

Let us represent the unit-radius $S^5$ metric as that of $S^1\cross S^3$ fibered over an interval $\rho\in[0,\frac{\pi}{2}]$ 
\begin{equation}\label{Smetric}
\dd s_{S^5}^2=\dd \rho^2+\cos^2\rho \,\dd s_{S^1}^2+ \sin^2\rho\,\dd s^2_{S^3}=\dd \rho^2+\cos^2\rho \,\dd \chi^2+ \sin^2\rho\Big(\sigma_x^2+\sigma_y^2+\sigma_z^2\Big)\,.
\end{equation}
Here we parametrised $S^1$ by $\chi\in[0,2\pi]$ and $S^3$ 
as in \eqref{sigma} with $\psi\in[0,4\pi]$. 
 A schematic picture of this parametrisation is given in figure \ref{figure1}a.
At $\rho=0$ (the ``north pole") 
 $S^3$ shrinks to a point and we recover a local 
 $\mathbb{R}^4\cross S^1$ geometry. 
 At the ``south pole" $\rho=\frac{\pi}{2}$ where $S^1$ shrinks to a point the local geometry is 
 $\mathbb{R}^2\cross S^3$. 

 To get the metric of the $S^5/\mathbb{Z}_2$ orbifold we restrict $\psi$ to $[0,2\pi]$. This space is then singular at $\rho=0$ with 
 $S^1_\chi$ at the north pole being the fixed circle. Thus,
 $S^5/\mathbb{Z}_2$ looks like $S^1\cross S^3/\mathbb{Z}_2$ fibered over the $\rho$-interval.

Motivated by the comparison of the $S^3$ parts of \rf{sigma} and \eqref{Smetric}, and by 
the Eguchi-Hanson resolution \eqref{244} 
of the $\mathbb{C}^2/\mathbb{Z}_2$ orbifold let us consider the following Ansatz for a resolution $\M^5$ 
 of $S^5/\mathbb{Z}_2$
\begin{equation}\la{28}
ds^2_{\mathcal{M}^5}= V(\rho)^{-1}\dd \rho^2 + \cos^2\rho \,\dd \tilde\chi^2+ \sin^2\rho\, \Big[\sigma_x^2+\sigma_y^2+ V(\rho)\,\tilde\sigma_z^2\Big]\, . 
\end{equation}
Here we ``deformed'' \eqref{Smetric} by introducing a function $V(\rho)$ and defined 
\begin{equation}\label{modforms}
\tilde{\chi}= \A\, \chi\,,\qquad\qquad \tilde\sigma_z= \B\, \dd \psi + \cos \theta\, \dd\phi \ , 
\end{equation} 
where $\A$ and $\B$ are two constants which are to be fixed momentarily. 
To provide a resolution of the orbifold singularity, 
the function $V(\rho)$ should behave as the Eguchi-Hanson one \eqref{244} near $\rho\sim0$, i.e.
 $V(\rho)\sim 1-(\tfrac{a}{\rho})^4$, and it should approach a constant for $\rho\sim \frac{\pi}{2}$.
 This is accomplished by the following choice 
\begin{equation}\la{210}
V(\rho)=1-\Big(\frac{\sin{a}}{\sin{\rho}}\Big)^4\,\ ,\qquad\qquad a\in \Big(0,\frac{\pi}{2}\Big)\,.
\end{equation}
Remarkably, \rf{28} with \rf{210} satisfies the same 5d Einstein equation as the undeformed metric \rf{Smetric}:
\begin{equation}
\mathcal{R}_{ab}=4g_{ab}\,, \qquad \qquad \mathcal{R}=20\,.\la{211}
\end{equation}
Let us note that although we can use this metric \rf{28} as a base for a Ricci-flat 6d cone, there is no associated K\"ahler form, so this resolution is not a Sasaki-Einstein space and breaks supersymmetry completely.

Let us first investigate the behaviour of $\M^5$ at $\rho=a$ where it appears to be singular as $V(a)=0$.
 Applying the following coordinate transformation from $\r$ to $u$ (cf. \rf{25}) 
\begin{equation}
u^2=\sin^2 \rho~ V(\rho) \ , \la{212}
\end{equation}
implies that for $u \to 0$ (or $\rho \to a$) the metric becomes
\begin{equation}
\dd s_{\mathcal{M}^5}^2\Big|_{u\to 0}\ \to\ \frac{1}{4} \Big[ \frac{1}{\cos^2 a} 
\, \dd u^2 + 4\cos^2 a\, \dd \tilde\chi^2+ 
{\sin^2 a}\,\dd s^2_{S^2}+ { u^2}(\B\, \dd \psi + \cos \theta\, \dd \phi)^2\Big]\,. \la{213}
\end{equation}
For this to be smooth at $u=0$ we have to choose
\begin{equation}
\B=({\cos a})^{-1} \,.\la{214}
\end{equation}
In this case the original orbifold singularity at $\rho=0$ is removed like in the above EH example. 

This comes at the cost of deforming the $S^3/\mathbb{Z}_2$  (parametrised by $\phi,\theta,\psi$)
 into an irregular 3-space.\footnote{We 
 thank J. Gauntlett for pointing out this issue missed in the original version of this paper.} The coordinate singularities at $\theta=0,\pi$ can be resolved in two separate coordinate patches. However, for generic values of $a$ these coordinate patches cannot be glued together consistently. 
Indeed, for $\tilde\sigma_z\sim \dd \psi+ q^{-1}   \cos \theta \,\dd \phi$ to be a well-defined fibre over $S^2$, we have to require that
\begin{equation}
{1\ov 2 \pi} \int_{S^2} \dd\left(q^{-1}  \cos \theta\, \dd \phi\right)=- 2 q^{-1} \in \mathbb{Z}
\end{equation}
is an integer (which then corresponds to a Lens-space).  
 For $ q^{-1} =\cos a $   this only holds 
 for the special values $a=0,\tfrac{\pi}{3} $ and $\tfrac{\pi}{2}$, which can be identified, respectively, as the 
 original orbifold case, a regular $S^3$ fibration \cite{Hashimoto:2004kc,Lu:2004ya}\footnote{This $S^3$ is fibred non-trivially over the other $S^2$ space and thus topologically distinct from $S^3\cross S^2$.} and the $S^3\cross S^2$ limit to be discussed below. 
For the purposes of this paper we will only consider quantities which are independent of the $\psi$-fibration and are therefore insensitive to these singularities. 
Additionally, we will take the orbifold limit $a\to0$ in the end, where the  
above 3-space 
 becomes regular (but we recover the orbifold singularity).

To study the vicinity of the ``south pole" at $\rho=\frac{\pi}{2}$ let us introduce the coordinate 
 $v=\frac{\pi}{2}-\rho$ so that the metric becomes 
\begin{equation}\label{oppole}
ds_{\mathcal{M}^5}^2\Big|_{\rho \to \frac{\pi}{2}} \to 
 \frac{1}{1-\sin^4a} \dd v^2 + v^2\A^2\, \dd\chi^2 + \frac{1}{4}\Big[\,\dd s_{S^2}^2+ ({1-\sin^4a})\Big(\frac{1}{\cos a}\dd \psi + \cos \theta\, \dd \phi\Big)^2\Big] \,.
\end{equation} 
For this to be smooth at $v=0$ we need to choose
\begin{equation}
\A=(1-\sin^4a)^{-1/2} \,.\la{216} 
\end{equation}
Note that both $q$ \rf{214} and $p$ become 1 at $a=0$.

To summarise, using \rf{28},\rf{210},\rf{214},\rf{216} we thus find a  
 ``resolution"
$\mathcal{M}^5$ 
of $S^5/\ZZ_2$ which is an Einstein space with the metric 
\ba 
\label{xxx}
&\dd s_{\mathcal{M}^5}^2=V(\rho)^{-1}\,\dd \rho^2+\frac{\cos^2\rho}{1-\sin^4 a} \,\dd \chi^2+\frac{1}{4}\sin^2\rho\, \Big[ \,\dd s_{S^2}^2+ V(\rho)\, \Big(\frac{1}{\cos a}\dd \psi + \cos \theta\, \dd \phi\Big)^2\Big] \,, \\
& \dd s_{S^2}^2=\dd \theta^2 +\sin^2 \theta\,\dd\phi^2\ , \quad 
 \rho\in[a,\frac{\pi}{2}]\,, \ \ \chi \in [0,2\pi]\,, \ \ \theta\in[0,\pi]\,, \ \ \phi\in[0,2\pi]\,, \ \ \psi\in[0,2\pi] \,. \no 
\ea
For generic values of $a$ it is singular at $\theta=0$ or $\pi$, but the 
original 
orbifold singularity has been regularised.  
 As an illustration of this resolution let us note that the square of  the  curvature of \rf{xxx}  is given by 
\begin{equation}\la{2.18}
\mathcal{R}_{abcd}\mathcal{R}^{abcd}=40+24 \frac{\sin^8 a}{\sin^8\rho}\Big(3+16\frac{\cos^2\rho}{\sin^4\rho}\Big)\,,
\end{equation}
which is finite for $\rho\in(a,\frac{\pi}{2}]$. 
If we consider the limit $\rho\to a$ and then $a\to 0$ we recover the orbifold singularity as 
$\mathcal{R}_{abcd}\mathcal{R}^{abcd} \to \ 384\, a^{-4}$. 

The volume form and the volume of $\M^5$ are given by
\begin{equation}\begin{split}\label{volume}
\vol_{\mathcal{M}^5}&=\frac{\sqrt{1+\sin^2 a}}{8\,(1-\sin^4 a)}\sin^3\rho \cos\rho\sin\theta\,\dd\rho \wedge \dd\chi\wedge\dd\theta\wedge\dd\phi\wedge\dd \psi\,,\\
\VV_{\M^5}&=\int_{\mathcal{M}^5}\vol_{\mathcal{M}^5}=\frac{\pi^3}{2}\sqrt{1+\sin^2 a}\,.
\end{split}\end{equation}
In the limit $a\to 0$ we recover the volume of the orbifold $\VV_{S^5/\ZZ_2} = \ha \VV_{S^5} = \frac{\pi^3}{2}$.

Let us comment also on the limit $a\to \frac{\pi}{2}$. 
In terms of the coordinate $v=\frac{\pi}{2}-\rho\in[0, b]$ in \eqref{oppole} where $b=\frac{\pi}{2}-a$ we 
can write \rf{xxx} as 
\begin{equation}
\dd s_{\mathcal{M}^5}^2= V(v)^{-1}\dd v^2+ \frac{\sin^2 v}{1-\cos^4 b}\,\dd \chi^2+\frac{1}{4} \cos^2v\Big[
 \,\dd s_{S^2}^2+ 4 V(v)\tilde\sigma_z^2\Big]\,,\quad V(v)=1-\big(\frac{\cos{b}}{\cos v}\big)^4.\ \ \ \ 
\end{equation}
 Setting $v= b \cos\eta$ and then taking the limit $b\to 0$ we get 
\begin{equation}
\dd s_{\mathcal{M}^5}^2\Big|_{a=\frac{\pi}{2}} = {1\ov 4}\Big[2 \dd \eta^2+ 2\cos^2 \eta\, \dd \chi^2+ \dd s_{S^2}^2 
+\,2 \sin^2\eta\, \dd \psi^2\Big]\,, 
\end{equation}
which is the metric of $S^3\cross S^2$. Thus 
 for $a$ changing from $0$ to $\pi\ov 2$ the metric \eqref{xxx} of $\M^5$ interpolates between $S^5/\mathbb{Z}_2$ and $S^3\cross S^2$ 
 (cf. figure \ref{figure1}b). 

In view of \rf{211}, we conclude that the resolved space $AdS_5 \times \M^5$ 
gives a solution of type IIB supergravity if we supplement it by a direct generalisation of the standard RR 5-form 
\begin{equation}
F_5= 4 \,(\vol_{AdS_5}+ \vol_{{\cal M}^5})\,. \la{222}
\end{equation}

\section{Twisted sector modes from 2-form fields in $AdS_5 \times \M^5$}\label{sec 3}

As discussed in the introduction, the twisted sector operator $T_k$ in \rf{ops} 
 should be dual to a 
 scalar mode in \adsz localised on $AdS_5\cross S^1$ and having $k$ units of momentum along the $S^1$. 
 Motivated by the idea of identifying 
 twisted sector fields using a blow-up procedure 
 of orbifold singularities \cite{Douglas:1996sw},
 we expect this 
 mode to originate from a combination of fluctuations of $B_2$ and $C_2$ fields in \rf{0} 
 that are ``wrapping" the blow-up 2-cycle
 in the resolved space $AdS_5 \times \M^5$ (cf. \rf{17}).

To justify this picture, we first study the type IIB supergravity 
equations for fluctuations of $B_2$ and $C_2$ fields 
in the $AdS_5 \times \M^5$ background. 
We identify modes which appear due to the resolution and 
correspond to the lightest twisted sector states in the orbifold limit $a\to 0$.
We then discuss the construction of their 6d effective action, 
which should reproduce 
the action \rf{18} of \cite{Gukov:1998kk}. 

\subsection{Solution of supergravity equations for 2-form fields}\label{sec 3.1}

The type IIB supergravity equations for the $B_2$ and $C_2$ fields following from \rf{0} may be written in
equivalent real and complex forms as 
\begin{align}
\dd \sstar F_3 -F_5\wedge H_3=0\,,\qquad \qquad 
\dd \sstar H_3 + F_5\wedge F_3=0\ ,
 \la{31}
\end{align}
\begin{equation}
\dd \sstar G_3 + i F_5\wedge G_3 =0 \ , \qquad \qquad G_{3}\equiv F_3 + i H_3=\dd A_2\,, \qquad\qquad A_2 = C_2+i B_2\ . \la{32} 
\end{equation}
We study these equations 
 in the background of the $AdS_5 \times \M^5$ metric \eqref{xxx} (of radius $L=1$) supported by the 5-form flux given in \eqref{222}.

Assuming 5+5 separation of coordinates,
 let us choose the following Ansatz for the potential $A_2$ in \rf{32}\foot{We use coordinates $x^i$ with indices $(i,j,...)$ for $AdS_5$
 and 
$y^a$ with indices $(a,b,...)$ for $\mathcal{M}^5$. 
In general, we
 shall use capital latin indices for coordinates of a generic 10d spacetime. Small latin indices from the beginning of the alphabet ($a,b,c,...$) label coordinates of internal 5-space 
and latin indices from the middle of the alphabet ($i,j,k,...$) label coordinates of non-compact 5-space 
or $AdS_5$.} 
\begin{equation}
 A_2(x,y)=\bbeta(x)\, \Omega(y)\,, \la{34}
\end{equation}
where $\Omega(y)$ is a complex 2-form on $\M^5$ and $\bbeta(x)$ is a real scalar function solving a
 free massive scalar equation in $AdS_5$
\begin{equation}
{\ssp}\bbeta=m^2 \bbeta\,.\la{35}
\end{equation}
Then \rf{31} is solved if $\Omega(y)$ on $\M^5$ satisfies\foot{Here and below 
$\star$ stands for the Hodge-dual 
 form in $\mathcal{M}^5$.}  
\be
\dd \star \Omega =0\,,\qquad \qquad 
\dd\star \dd \Omega - 4i\dd\Omega + m^2 \star \Omega=0\,. \la{36}
\ee
These equations are satisfied if we express $\Omega$ in terms of a closed 3-form $\omega$ as 
\be
\Omega=\star\, \omega\,,\qquad \qquad \dd\omega=0\,, \qquad \qquad 
\Big[(\dd\star)^2 -4 i(\dd \star) + m^2 \Big]\omega =0\,. \la{37}
\ee
Equivalently, the closed 3-form $\omega$ on $\M^5$ should satisfy
\begin{equation}\label{onset}
\dd \star \omega = -i M \omega\ , \qquad \qquad 
M (M + 4)=m^2\,.
\end{equation}
The complex conjugate field $\bar{A}_2={\varphi}(x)\bar\Omega(y)$ 
should solve  
the complex conjugate equations, so that $\bar\Omega(y)$ is expressed in terms of $\bar \omega$ as (we assume that $m$ is real)
\begin{equation}\label{spectrum2}
\dd \star \bar\omega = -i M \bar\omega\,, \qquad\qquad 
 M (M - 4)= m^2 \,.
\end{equation}
The metric 
\rf{xxx} of $\M^5$ has an isometry 
along the $S^1$ parametrised by $\chi$ so we may expand in Fourier modes 
\ba 
\omega(y)=\sum_{k=-\infty}^\infty e^{i k \chi}\, \omega_k(\rho,\theta,\phi,\psi)\,,\la{310}
\ea 
and, for the time being, focus on one particular mode $\omega_k$ with positive $k$.\footnote{
Note that 
 $\omega$ and $\bar{\omega}$ solve the same differential equation \eqref{onset}, so we may consider only $\omega_k$ with $k>0$ and relegate $k<0$ to modes of $\bar{\omega}$. 
}

\begin{figure}[h]
	\centering
	\includegraphics[width=0.5\textwidth]{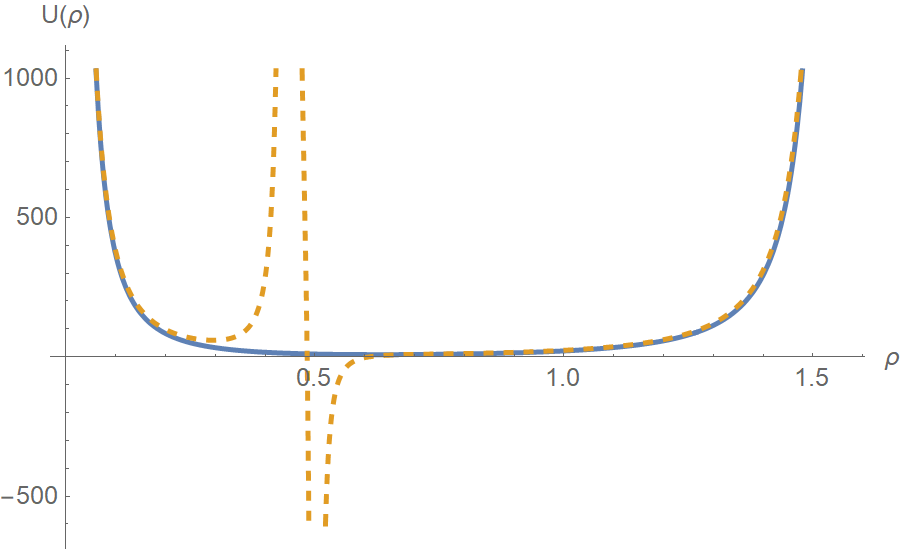}
	\caption{\small{Potential $U(\rho)$ for $k=-M=3$ and (i) $a=0$ (solid curve) 
	and (ii) $a=\frac{1}{2}$ (dotted curve). 
	}
	}\label{figure2}
\end{figure}

As we are interested in the analogues of the 
twisted sector modes, which are to be localised near the north pole $\r=0$
of the orbifold we may restrict our attention to the lowest harmonics on the deformed $S^3$ part of \rf{xxx}.
Explicitly, we may assume no dependence on $\psi$ and a spherical symmetry in $(\theta,\phi)$.
A general Ansatz for such $\omega$ then takes the form\footnote{We do not include
$f_0(\rho)\,\sigma_x\wedge\sigma_y\wedge\tilde\sigma_z$ as this form is not closed. $\tilde{\sigma}_z$ is not well defined at $\theta=0$ and $\pi$,
as was pointed out above.} 
\begin{equation}\label{311}
\omega_k =f_1(\rho)\, \dd\rho\wedge\dd\tilde\chi\wedge\tilde\sigma_z
+f_2(\rho)\, \dd\rho\wedge\sigma_x\wedge\sigma_y+f_3(\rho)\, \dd\tilde\chi\wedge\sigma_x\wedge\sigma_y\,,
\end{equation}
where $\dd\tilde\chi$ and $\tilde\sigma_z $ were defined in \eqref{modforms},\rf{214},\rf{216}.
Then the equation \rf{onset} relates $f_1$ and $f_2$ to $f_3 $
with the latter being subject to a 2nd order ordinary differential equation
(see appendix \ref{appendix C}). 
This equation can be put into the Schr\"odinger-type form
\begin{equation}\label{312}
\gf''(\rho)-U(\rho) \ \gf (\rho)=0\,, 
\end{equation}
where $\gf$ is related to $f_3$ via 
rescaling by a function of $\rho$ (see \rf{c11}).
 The potential $U(\rho)$ 
 is depicted in figure \ref{figure2} for some special values of the parameters. 

At $a=0$ the potential has the form of a well between two second-order poles
 and the equation \rf{312} can be solved explicitly, generating a discrete spectrum of solutions. 
 In the case of 
 \ba M = - k \ , \la{333} 
\ea
 we find for the 2-form $\Omega$ in \rf{37},\rf{310},\rf{311}
\begin{equation}
\label{313} \Omega = e^{ik\chi}\, \Omega_k \ , \quad 
\Omega_k= \frac{\cos^k\rho}{\sin^2\rho}\Big[(2+k\sin^2\rho)\, \sigma_x\wedge\sigma_y
 + ik\sin^2\rho\,\dd\chi\wedge\sigma_z -\frac{2\cos^2\rho+k\sin^2\rho}{\cos\rho\sin\rho}\, \dd\rho\wedge\sigma_z\Big]\,.\end{equation}
This solution is not normalisable as it diverges near $\rho \to 0$.
It does not have an analogue in the 
 KK spectrum \ci{Kim:1985ez} of the usual $S^5$ compactification, 
 which starts with the first normalisable solution at $M=k+2$ (see appendix \ref{appendix C}). 

\begin{figure}[h]
	\centering
	\includegraphics[width=0.5\textwidth]{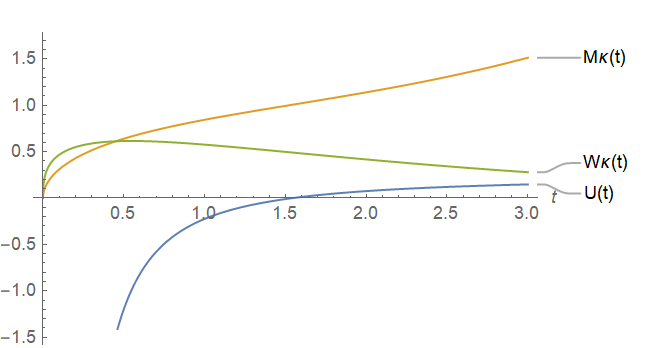}
	\caption{\small{Whittaker equation potential and solutions for small $a$.}}\label{figure3}
\end{figure}

For $a>0$ the potential $ U(\rho) $ develops a pole at $\rho=a$ (cf. figure \ref{figure2}):
\begin{equation}\la{314}
U(a+\epsilon)=-\frac{1}{4\epsilon^2}-\frac{\kappa\mu}{\epsilon}+\frac{\mu^2}{4}+\order{\epsilon}\,, \qquad 
\kappa\sim \frac{1}{2\sqrt{5}}+\order{a^2}\,,\qquad\mu\sim\frac{16\sqrt{5}}{k^2a^3}+\order{a^{-1}}\,, 
\end{equation}
where we gave the small-$a$ expansions of the coefficient functions $\kappa(a,k,M)$ and $\mu(a,k,M)$ 
entering the expression for $U(\rho)$. 
 Near $\rho =a$ the equation \eqref{312} can be transformed into the standard Whittaker equation form 
by a rescaling $\epsilon=\rho-a\to\mu^{-1}{t}$ (here the derivatives are w.r.t. $t$) 
\begin{equation}\la{315}
\gf''(t)+\Big(-\frac{1}{4}+\frac{\kappa}{t}+\frac{1}{4t^2}\Big)\gf(t)=0\,.
\end{equation}
This equation 
is solved by the Whittaker functions ${\rm M}_{\kappa,0}(t)$ and ${\rm W}_{\kappa,0}(t)$,
 which may be expressed in terms of confluent hypergeometric functions.
 We plot these functions and the potential $U(t)$ of \rf{315}
 in figure \ref{figure3}\,.
 Both functions vanish asymptotically at $\rho\to a$, i.e. the resolution provides a regularisation of the $\r=0$ singularity 
 present in the orbifold limit. 

In general, as the potential $U(\rho)$ in \rf{312} is smooth, there are two solutions
of \rf{312}. Close to $\r=a$ they look like the Whittaker functions,
 while away from this point they look like the solutions found in the $a=0$ case (cf. \rf{313}). 
 The full solution may be constructed numerically. 
 
 This implies that a full 
 solution generalizing the one in \rf{313} found for $a=0$ 
 (which was divergent at $\r=0$ and thus potentially discarded as non-normalisable)
 is regular and normalisable for $a\not=0$. 
 In the limit $a\to 0$ such solutions appear to be localised near $\r=0$. 
 This suggests that taking the limit $a\to0$ we need to keep these modes in the spectrum, and they 
 should represent the light 
 twisted sector states.

\subsection{Effective action for twisted sector modes} \label{sec 3.3}

The localised modes of 2-form fields discussed above propagate on the $AdS_5\cross S^1$ subspace at the north pole 
 of $AdS_5\cross \M^5$. 
 Expanding in modes of $\M^5$ in general yields a KK tower of massive fields in $AdS_5$ (cf. \rf{35}). 
 We focus on the twisted sector solutions corresponding to 
 \rf{313} 
 for which 
\begin{equation}\label{399}
 M=\pm k \ , \qquad \qquad m^2=k^2\pm 4k \ , 
\end{equation}
 where $k$ is the $S^1$ mode number. 
 As the ``transverse'' part $\Omega(y) $ of the field $A_2= C_2 + i B_2$ in \rf{34} 
 depends on $k$ 
 we label the corresponding $AdS_5$ part as $\bbeta_k(x)$.

We may then reconstruct the corresponding 5d effective action for $\bbeta_k(x)$ by starting with 
the 10d supergravity action in \rf{0}. Written in terms of $G_3$ defined in \rf{32}
the relevant part 
reads 
\begin{equation}\label{777}
S_{10}=-\frac{1}{4\kappa^2}\int \Big[\bar{G_3}\wedge \sstar G_3 + i \bar A_2 \wedge G_3\wedge F_5\Big]\,.
\end{equation}
Inserting here the Ansatz \rf{34} with $\Omega$ given by \rf{37},\rf{onset} and integrating over $\M^5$
we get the effective action for $\bbeta_k$ propagating in $AdS_5$ 
\ba\label{317}
S_{5}\ \sim& 
\ \ha \mathcal{V} \int\dd ^{5} x \sqrt{-g_{_{AdS_5}}}\ \bbeta_k \, \big({\ssp}-m^2 \big)
\, \bbeta_k\ , \qquad m^2= k(k-4)\ , \\ 
& \qquad \mathcal{V}\equiv \int_{\mathcal{M}^5}\bar\omega_k\wedge \star\omega_k\,. \la{319}
\ea
For every real $\varphi_k$ there is also mode $\varphi_{-k}$ arising from $\bar{\omega}$ \eqref{spectrum2}
with $m^2 = k ( k+4)$. 
For the Ansatz \eqref{311} the prefactor \rf{319} becomes
\begin{equation}\begin{split}\label{norm}
\mathcal{V}&=\tfrac{1}{8}\int\sin\theta\, \dd\chi\dd\theta\dd\phi\dd\psi \int_a^{\frac{\pi}{2}}\dd \rho\,\frac{\sin^2\rho \abs{f_1(\rho)}^2+\cos^2\rho\, V(\rho)\abs{f_2(\rho)}^2+ \abs{f_3(\rho)}^2}{\cos\rho\sin\rho}\,.
\end{split}\end{equation}
Inserting 
the solution \rf{313} corresponding to the twisted sector mode of interest 
we find for $a\to 0$
\begin{equation}\la{322}
\mathcal{V}\sim \frac{4\pi^3}{a^4}+\order{a^{-2}}\,.
\end{equation}
 This suggests that 
 we should rescale 
 $\vp_k$ by $a^2$, getting $\mathcal{V} \to a^4 \mathcal{V}$ and thus 
 a finite action in the $a\to 0$ limit.\foot{\la{f11}With this $a^4$ factor added,
 the integrand of the $\rho$-integral in the region where 
 $ \rho \to a \to 0$ takes the form of $\delta(\rho-a)$ 
 restricting the integration to a small $S^3/\mathbb{Z}_2$-shell around $\rho=0$. 
 This may be interpreted as the orbifold singularity being cut off at scale $a$. 
 }

Alternatively, instead of starting with the 5+5 Ansatz \rf{34} 
we may follow the idea of \cite{Gukov:1998kk} and attempt to construct an 
 effective 6d action for twisted modes localised 
on $AdS_5\cross S^1$. 
Locally, near $\r=0$, 
 we may approximate $S^5/\mathbb{Z}_2$ as $ \mathbb{C}^2/\mathbb{Z}_2\cross S^1$, so that 
 the $S^1$ dependence factorises. 
One may try to justify this approach by using the solution on the resolved manifold $\mathcal{M}^5$ 
 found above.
The solution for the 2-form $\Omega$ in \rf{34} given by \rf{313} is regular everywhere apart from $\rho=0$, where 
 \begin{equation}\la{327}
 \Omega\ \overset{\rho\to0}{\sim}\ \frac{2 e^{ik\chi}}{\sin^2\rho}\Big(\sigma_x\wedge\sigma_y-\frac{\dd(\sin\rho)}{\sin\rho}\wedge \sigma_z \Big)\ \to \ e^{ik\chi}\, \Theta (\rho,\theta,\phi,\psi) \ , \ \qquad \ \ \ \ 
 \Theta\equiv \dd\Big(\frac{1}{\sin^2\rho} \sigma_z\Big)\ . 
 \end{equation}
In the vicinity of $\r=0$ we thus find that $\Omega$ is an exact 3-form up to an $e^{ik\chi}$ phase factor. 
This suggests that for $a\to 0$ (when the relevant modes localise close to $\rho=0$) 
we may effectively decouple the $S^1_\chi $ from the rest of $\mathcal{M}^5$. 
 Then starting with the factorised form of the field $A_2$ in \eqref{34} 
 and summing over positive and negative $k$-modes
 we find that the part of $A_2$ that is divergent for $\rho \to 0$ 
 factorises as
 \ba\la{327a}
 A_2= &\sum_{k=-\infty}^{\infty}\varphi_k(x) e^{ik\chi} \Omega_k (\rho,\theta,\phi,\psi)\ \overset{\rho\to0}{\sim}\ 
 \hat \varphi(x,\chi)\, \Theta (\rho,\theta,\phi,\psi)\,, \\
 \hat \varphi(x,\chi)=& 
 \sum_{k=-\infty}^{\infty}e^{ik\chi} \, \varphi_k (x) \ . \la{1111}
 \ea
Assuming that this divergent part 
which is regularised by the blow-up procedure 
dominates over contributions coming 
from the regular part of $\Omega_k$ in \eqref{313} we may thus approximate our 2-form fields as in \rf{17}
by 
 \begin{equation}\la{323} 
B_2=\beta(x,\chi) \,\Theta\,,\qquad\qquad C_2=\gamma(x,\chi) \,\Theta \ , \ \ \ \ \ \ \ \ \ \ 
\hat \vp (x, \chi) = \beta(x,\chi) + i \gamma (x,\chi) \ , 
\end{equation} 
where the real 2-form $\Theta$ is independent of the $S^1$ coordinate $\chi$. 
The resulting action for the 6d fields 
$\beta$ and $\gamma$ following from \rf{0} or \rf{777} has the following structure 
\ba\label{325}
&S_6\ \sim \ 
\ha
\int_{AdS_5\cross S^1}\dd^5 x\ {\dd\chi} \,\sqrt{-g_6}\,\Big[ {c_1}(\partial_i\beta)^2+ {c_2} (\partial_\chi\beta)^2+
{c_1} (\partial_i \gamma)^2+ {c_2} (\partial_\chi\gamma)^2 -8 c_3 \, \beta\,\partial_\chi \,\gamma \Big]\,,
\\
\label{888}
&c_1=\pi\int 
\dd^4 y\sqrt{g}g^{ac}g^{bd}\Theta_{ab}\Theta_{cd}\,,\ \ 
c_2 =\pi \int
\dd^4 y\sqrt{g}g^{ac}g^{bd}\Theta_{ab}\Theta_{cd} g^{\chi\chi}\,,\ \ 
c_3=-2 \pi \int
\Theta\wedge\Theta\,, \ea
where the integrals go over the local factor-space $\M^5/S^1$. 

In the case of the flat-space orbifold we may choose an anti-self-dual 
$\Theta$ and $g^{\chi\chi}=1$ so that 
 $c_1=c_2=c_3$. 
 In the $S^5/\ZZ_2$ orbifold case using
 the approximation \eqref{323} and the definition of $\Theta$ in \rf{327} as well as \rf{28},\foot{Explicitly, one has 
\begin{equation*} 
\frac{1}{2}g^{ac}g^{bd}\Theta_{ab}\Theta_{cd}=\frac{8}{\sin^{8}\rho},\quad 
g^{\chi\chi}=\frac{1}{\cos^2\rho }\,, \quad
\sqrt{g}=\frac{1}{8}\sin^3\rho \cos\rho\sin\theta\,.
\end{equation*}}
we find that integrands of $c_1,c_2$ and $c_3$ in \eqref{888} diverge 
as $\rho^{-5}$ for $\rho \to a\to 0$ implying that
\begin{equation}
c_1\sim c_2\sim c_3\sim \frac{4\pi^3}{a^4}+\order{a^{-3}}\,.\la{330}
\end{equation}
Notice that this divergence is of the same order as in \eqref{322} and should be treated similarly. 
As a result, we get essentially 
 the same expression as in \rf{18} for the leading contribution to the 6d effective action
\begin{equation}\label{6d}
S_6\sim 
\ha \mathcal{V}
\int_{AdS_5\cross S^1}\dd^5 x\, {\dd\chi}\,\sqrt{-g_6}\,\Big[(\partial_i\beta)^2+(\partial_\chi\beta)^2+(\partial_i \gamma)^2+(\partial_\chi\gamma)^2-8\, \beta\,\partial_\chi\gamma \Big]\,.
\end{equation}
 Expanding $\beta$ and $\gamma$ in Fourier modes along $\chi$ 
 and diagonalising the action (cf. \rf{19}) 
 results in the same spectrum and 5d action as in \eqref{399},\rf{317}.

\section{Matching to gauge theory: leading and subleading corrections}\label{sec 4}

Let us now address the question of reproducing 
the gauge theory prediction for the 2-point correlator of twisted sector modes in \rf{3}. 
We shall first comment on the matching at the leading order in strong coupling \cite{Billo:2022fnb}
 based on the low-energy action \rf{18} 
 and then discuss how to modify this action to include leading $\a'^3$-corrections.

\subsection{Normalisation factors}\label{sec 4.1}

Let us start with an action for a massive scalar in $AdS_5$
\begin{equation}\la{41}
S[\bbeta]=\frac{1}{2}\Vv_{\vp} \int \dd^5x\sqrt{-g_{_{AdS_5}}}\ \bbeta\, \big({\ssp}-m^2 \big) \bbeta\,, \qquad \ \ \Delta(\Delta-4)=m^2\,, 
\end{equation}
where for generality we introduced a normalisation factor $\Vv_\vp$.
Fixing the Dirichlet boundary condition $\vp|_{_{\partial AdS}} = \vp_0$ one gets for the value of the action 
 \cite{Witten:1998qj,Freedman:1998tz}
\begin{equation}\la{42}
S[\varphi_0]=\frac{1}{2}\Vv_\vp \, \NN_\Delta \int_{\partial AdS_5}\dd^4x\dd^4x'\ \frac{\varphi_0(x)\varphi_0(x')}{\abs{x-x'}^{2\Delta}}\,,\qquad \qquad 
\NN_\Delta=\frac{2}{\pi^2}\frac{\Delta-2}{\Delta}\frac{\Gamma(\Delta+1)}{\Gamma(\Delta-2)}\,.
\end{equation}
We assume 
that the generating functional for the correlators of the corresponding 
dual operator $\mathcal{O}$ contains the source term 
$
\Ca_\mathcal{O}\int_{\partial AdS_5}\varphi_0 \mathcal{O}\,. 
$
The resulting prefactor in the 2-point function of $\mathcal{O}$ is given by (cf. \rf{1.1}) 
\be\la{443} 
{\rm G}_\mathcal{O}= \, \NN_\Delta\, \Vv_\vp\, \Ca_\mathcal{O}^2 \,.
\ee
In general, the value  of ${\rm G}_\mathcal{O}$ is 
ambiguous depending 
on normalisations of the field $\varphi$ 
and the dual operator $\OO$. 

In ref. \cite{Billo:2021rdb} the 
 dimensionless action normalisation constants $\Vv_\vp$ in the 5d actions corresponding to 
 the untwisted and twisted scalar fields dual to the operators $U_k$ and $T_k$ in \rf{ops} 
 were given as\footnote{Here $L$ is the scale of $AdS_5$ and 
 $ \frac{1}{2\kappa^2}=\frac{1}{(2\pi)^{7}\alpha'^4g_s^2}=\frac{4(2N)^2}{(2\pi)^5L^8}\,,\quad s_k= 2^{6-k}\frac{k(k-1)}{(k+1)^2}\,,
$ where $2N$ is the rank of the gauge group before orbifolding.}
\begin{align}
\Vv_{\vp_{_{U_k}}}&=\frac{1}{2\kappa^2}\cdot L^3 \cdot \frac{\pi^3}{2}L^5\cdot s_k=\frac{N^2}{2^{k-4}\pi^2}\frac{k(k-1)}{(k+1)^2}\,,\la{47}\\
\Vv_{\vp_{_{T_k}}}&= \frac{1}{2\kappa^2} (\pi\alpha')^2 \cdot L^3\cdot 2\pi L=\frac{N^2}{\lambda\, \pi^2}\,.\la{48}
\end{align}
In \rf{47} the factor $\tfrac{\pi^3}{2}L^5$ is 
 the volume of $S^5/\mathbb{Z}_2$ and $s_k$ comes from the KK-mode overlap integral on the compact space
 (cf. \ci{Lee:1998bxa}). With the choice of normalisation \rf{ops} of $U_k$ 
 (corresponding to a particular 
 $\Ca_{\vp_{_{U_k}}}$ in \rf{443}) we have ${\rm G}_{U_k}=1$ in \rf{5}. 

The presence of the factor $(\pi\a')^2$ 
in the twisted field normalisation \rf{48} appears to be an ad hoc choice required to explain the presence of the non-trivial $\lambda^{-1}$ prefactor in ${\rm G}_{T_k}$ in \rf{3}.\foot{In \cite{Billo:2021rdb}, the factor $(2\pi\alpha')^2$ was introduces as a rescaling to make the boundary value of the field 
$\varphi_0$ dimensionless (with extra 4 to cancel $1/4$ factor in the 6d action there).}
It may be attributed to the ``stringy'' nature of the twisted sector modes described by a 6d action \rf{18} 
that should have an overall normalisation fixed directly from the string theory computation 
involving twisted-state vertex operators normalised in a particular way. It should be related to the 
localisation of the twisted-sector modes to the fixed 
6d subspace with the factor $\a'^2$ effectively replacing the 
``transverse'' 
4-volume 
factor $L^4$  in the untwisted case \rf{47}.

In our present approach where the starting point is the 10d supergravity action expanded near 
 the resolved orbifold background $AdS_5 \cross \M^5$ the role of this 
 extra scale factor is
 effectively played by the resolution parameter $a$. Indeed, as we discussed above, compactifying 
 from 10d to 5d we get the twisted-mode action \rf{317} with $a^{-4}$ scaling \rf{322}  
 representing the delta-function in 4 transverse directions (see also footnote \ref{f11})
 that can be eliminated by a rescaling $\vp \to a^2 \vp$. 
 Interpreting $a$ as an
 effective  counterpart of the string scale $\sqrt{\a'}$ in a direct string theory computation, this 
 produces an extra $({\a'\ov L^2})^2$ factor in the twisted case \rf{48} relative to the untwisted one \rf{47}. 

Going beyond the leading order in large $\l$ the form of the localisation result for ${\rm G}_{T_k}$
 in \rf{3} 
implies that one is to replace $\sqrt \l$ by $\sqrt{ \l'}$ according to \rf{44} \cite{Beccaria:2022ypy,Beccaria:2023kbl}. 
This redefinition 
 may be interpreted on the string side as being related to a renormalisation of the effective string tension 
 or of the $AdS_5$ radius which should be due to the fact that orbifolding breaks half of the maximal supersymmetry of $AdS_5 \cross S^5$, i.e.
 \be 
 \sqrt \l= {L^2\ov \a'}\ \ \to \ \ \sqrt{\l'} = {L^2\ov \a'}-4 \log 2 \ . \la{48a} 
 \ee
 The extra factor $\big(\frac{\lambda'}{\lambda}\big)^{k}$ in \rf{3} may be absorbed into the normalisation of the 
 twisted sector operators. 
 Up to the overall factor 
 the subleading corrections in the strong-coupling 
 expansion \rf{3} of ${\rm G}_{T_k}$ have the same pattern as $c_1\a'^3 + c_2 \a'^5+ ...$ corrections in 
 type IIB string theory. The first subleading 
 term  is 
\begin{equation}\la{411}
{\rm G}_{T_k}\sim 1+ \ha(2k-1) (2k-2)(2k-3) \frac{\zeta(3)}{\lambda'^{\frac{3}{2}}}+\OO({\lambda'^{-{5}/{2}}}) \ . 
\end{equation}
Let us now discuss a possible tree-level 
string theory origin of this large-$N$ strong coupling correction. 

\subsection{$\alpha'^3$-corrections}\label{sec 4.2}

Our strategy is to find a higher-derivative correction to the 6d action \rf{6d} quadratic in the twisted sector modes 
by starting with the tree level type IIB string effective action including $\a'^3 \zeta(3) \R^4+ ...$ terms 
and repeating the procedure that led from the 10d supergravity action to the action \rf{6d}. 

In section \ref{sec 3} we have shown that the fields $B_2$ and $C_2$ 
 develop additional normalisable modes on the resolved orbifold 
 background $\mathcal{M}^5$. Sending the resolution parameter $a$ to zero, we observed that these modes localise close to the emerging orbifold singularity and can be described by the effective 6d action 
 \rf{6d}.
 We now expand the relevant $\a'^3$-terms which are quadratic in $A_2= B_2 + i C_2$ 
 near the deformed orbifold background $AdS_5 \cross \mathcal{M}^5$ and
 use that the relevant modes localise on $AdS_5 \cross S^1$ to integrate over the internal 4-space.
 This should result in 
 $\a'^3$-corrections to the 6d action \rf{6d} 
 responsible for the subleading $\zeta(3)$-term in \rf{411}. 
 
To recall, the tree-level 
 type IIB string low-energy effective action has the following schematic form 
(see, e.g., \cite{Green:1982sw,Gross:1986iv,Sakai:1986bi})
\begin{equation}\la{415}
S_{\rm eff}= S_{10}+{\alpha'^3} \zeta(3)\int \dd^{10}x\,\sqrt{-g} \Big[ \mathcal {R}^4 + \mathcal{L}_8(\mathcal{R},F_5,G_3)\Big] +\order{\alpha'^5}\,.
\end{equation}
Here $S_{10}$ is the type IIB supergravity action, $ \mathcal {R}^4$ indicates the curvature-dependent 
invariant and 
 $\mathcal{L}_8$ depends on RR fields (we ignore numerical constants and dependence on 
 dilaton, RR scalar and fermions). Expanded near flat space, 
 $\mathcal{L}_8$ involves at least four fields and eight derivatives 
 as the 2-point and 3-point string amplitudes for the massless modes do not receive $\alpha'$-corrections.
 The explicit form of $\mathcal{L}_8$ 
 should be fixed by supersymmetry but is presently not known (cf. appendix \ref{appendix B})
 so our discussion below is partly qualitative. 
 
 Assuming a scheme (field redefinition) 
 choice in which the curvature dependence of the string effective action 
 is expressed in terms of 
 the 10d Weyl tensor $\C$ (i.e. replacing $\mathcal {R}^4$ in \rf{415} by $\C^4$, etc.) 
 one concludes that the 2-point and 3-point functions of the 
 massless string modes do also not receive corrections near 
 the conformally flat $AdS_5 \cross S^5$ background (cf. \ci{Banks:1998nr,Gubser:1998nz}). 
 The same then applies to the untwisted sector BPS modes in the $AdS_5\cross S^5/\mathbb{Z}_2$ orbifold case
 (the Weyl tensor here is again zero away from the orbifold singularity).
 
 To find the correction to the action \rf{6d} of the twisted sector modes we need to consider \rf{415} expanded near the resolved $AdS_5\cross \mathcal{M}^5$ background and determine terms quadratic in the $A_2$-field 
 that survive in the $a\to0$ limit.
 For $AdS_5\cross \mathcal{M}^5$ the Ricci tensor and $F_5$ have the same structure as for \adss (see \rf{211},\rf{222})
 \begin{equation}\la{4167}
\mathcal{R}_{ij}=-\frac{4}{L^2} g_{ij}\,,\qquad\qquad \mathcal{R}_{ab}=\frac{4}{L^2} g_{ab} \ , 
 \qquad \qquad F_5= \frac{4}{L} \big(\vol_{AdS_5}+\vol_{\M^5}\big)\ , 
\end{equation} 
 but 
 the Weyl tensor of $\M^5$ is no longer zero (cf. \rf{2.18},\rf{D8}) 
 \begin{equation}\la{410}
\mathcal{C}_{abcd}(\mathcal{M}^5)\ \sim\ \frac{\sin^4 a}{\sin^6 \rho}\,.
\end{equation}
 The leading correction to the term quadratic in $G_3=dA_2$ should then come 
 from the structures in $\mathcal{L}_8$ in \rf{415} that are at least linear in the Weyl tensor $\C(\M^5)$, i.e. 
 \begin{equation}\la{420}
\mathcal{L}_8\ \sim \ \mathcal{C} \, F_5\, \bar{G}_3\nabla^3 G_3 + \dots\,, 
\end{equation}
Here we indicated only the term with highest possible (third) power of covariant derivatives
as one can see on dimensional grounds. 
This term may be related to the $k^3$-term in \rf{411}. 
Its detailed index structure is discussed below. 

In general, assuming that the relevant modes of $B_2$ and $C_2$ ``localise'' to 6d space as in \rf{323} 
we are led to the following correction to the 6d action \rf{325} 
\ba\la{4122} 
& \Delta S_6\ \sim \ \alpha'^3\zeta(3)\, \Vv \, 
\int_{AdS_5\cross S^1}\dd^5 x\, {\dd\chi} 
\,\sqrt{-g_6}\,\Big(\beta\, \mathcal{K}_{1}\,\beta+ \gamma\, \mathcal{K}_{2}\,\gamma+\beta\, \mathcal{K}_{3}\,\gamma\Big)
\,,\\ \la{4222}
&\qquad \qquad \mathcal{K}_{r}= \sum_{n=0}^{2}
\sum_{l=0}^{5-2n} {\rm k}_{nl, r} \, ({\ssp})^n\ (\partial_\chi)^l \,.
\ea 
 The masses of the $\beta$ and $\gamma$ fields are expected to be protected by supersymmetry 
 so $\mathcal{K}_{1}$ and $\mathcal{K}_{2}$ should depend only on the 6d covariant combination 
 ${\ssp}+\partial_\chi^2$. 
The terms with second power of $\ssp$ in \rf{4222} can then be eliminated using field redefinitions (cf. \rf{325}) so we may ignore them. Furthermore, 
 the mixing of $\beta$ and $\gamma$ in \rf{4122} 
 may only affect normalisations at subleading order in $\alpha'$. 
 We may then assume that the only relevant effect of adding the correction 
 $\Delta S_6$ to the leading-order action \rf{6d} is a
 possible change of the overall normalisation due to an extra operator 
 $P(\del_\chi)= p_3 \del_\chi^3 + p_2 \del_\chi^2 + p_1 \del_\chi + p_0 $ factor in 
 $ \mathcal{K}_{r}$ operators, i.e. 
 \begin{equation}
\mathcal{K}_{1}= \mathcal{K}_{2}={P(\partial_\chi)}\Big({\ssp}+\partial_\chi^2\Big),\qquad \mathcal{O}_{\beta\gamma}=8 P(\partial_\chi)\, \partial_\chi\,.
\end{equation} 
After expanding in $S^1_\chi$ modes we have $\del_\chi \to i k$ so to reproduce the correction in \rf{411} we need 
\begin{equation}\la{424}
P(\partial_\chi)\ \sim \ \ha(2i \partial_\chi+1) (2i \partial_\chi+2)(2i \partial_\chi+3)\,.
\end{equation}
This peculiar structure should be dictated by supersymmetry. 

\section{Matching the $k^3$-term}\la{sec5} 
Let us now try to substantiate the above procedure and fix the required structure of the correction in \rf{420} by 
 focussing on the leading $\del_\chi^3$-term in \rf{424} that should reproduce the $k^3$-term in \rf{411}.
 
 This term is dominant in the large $R$-charge limit $k\to \infty$ 
 which, combined with the large-$\l$ expansion in \rf{3},  should be analogous to the familiar BMN limit \cite{Berenstein:2002jq}. 
 Indeed, the strong coupling expansion 
 of the gauge theory expression in \rf{3} expanded also at large $k$ 
 admits a regular limit 
\be
{\rm G}_{T_k}\Big|_{\lambda,\, k\to \infty} =(2\pi \nu)^2 \, e^{-8\nu\log 2} \Big[1
+ 4 \zeta(3)\, \nun^3 - 9 \zeta(5)\, \nun^5 + 8 \zeta(3)^2\, \nun^6 + ...\Big] \ , \ \ \ \ \ \ \ \ \ 
 \nun\equiv \frac{k}{\sqrt{\lambda} } \ . \la{5555} \ee
This gives a hint 
 that one may be able to reproduce \rf{5555} on the string theory side 
 by starting with a pp-wave 
 limit of the orbifold background.\foot{\la{f15}While this may work for the 2-point correlator, 
 in general, the pp-wave limit may not be enough for reproducing 3-point functions.
 Still, it is interesting to note that 
 according to \rf{5} and \rf{4} in the large-$k$ limit, we get 
 ${\rm G}_{U_k,U_k,\bar{U}_{2k}}\to \frac{k^{3/2} }{N} \ , \ \ 
 {\rm G}_{T_k,T_k,\bar{U}_{2k}}\to \frac{k^{{3}/{2}}}{N}\, (2\pi\nu)^2\, e^{-8\nu\log 2} 
 \big[1+ 4{\zeta(3)}\nu^3 + \OO(\nu^5) \big] $. Thus 
 the ratio ${\rm G}_{T_k,T_k,\bar{U}_{2k}}/ {\rm G}_{U_k,U_k,\bar{U}_{2k}}$ has a well-defined limit depending only on $\nu$. } 
 The parameter $\nun = {k\ov\sqrt \l}$ is the analogue of the semiclassical BMN-momentum along $S^1_\chi$ 
 which is fixed in the large-$k$, large-$\l$ limit. 
 
Below we suggest a strategy to reproduce 
 the $\nu^3$-term in \rf{5555} by starting with the pp-wave limit of the 
 resolved orbifold background $AdS_5\cross \M^5$. 
This provides a substantial simplification 
 allowing one to see more explicitly that $\alpha'^3$-corrections as in \rf{420} 
 indeed lead to the $k^3 \zeta(3)$-term
 in \rf{3},\rf{411} or the $\nu^3 \zeta(3)$-term in \rf{5555}. 
 
 At the end of this section we shall return to the case of the original $AdS_5\cross \M^5$
 and identify a particular structure in \rf{420} that may correspond to the $k^3$-term 
without first taking the pp-wave limit.

\subsection{Large-$k$ limit: pp-wave analogue of the resolved orbifold}\label{sec5.1}

In the familiar $AdS_5\cross S^5$ case the Penrose limit \ci{Blau:2001ne}
corresponds to focussing on states with a large momentum $k$ along $S^1\subset S^5$, i.e. expanding near a
 light-like geodesic along the time direction of $AdS_5$ and 
 an isometry circle of the $S^5$.
 This is equivalent to a scaling limit 
 \be \la{500}
 {L\ov \sqrt{\a'}}\to \infty\,, \ \ \, k\to\infty\,, \ \ \qquad \qquad 
 \nun=\tfrac{k}{\sqrt{\lambda}}=\tfrac{\a'}{L^2}k ={\rm fixed} \ . \ee 
In the present orbifold case we may consider a similar limit of $AdS_5\cross \mathcal{M}^5$ 
with $S^1$ being the fixed 
$\chi$-circle. 
Starting with 
\begin{equation}
\dd s_{10}^2 = L^2\Big(\dd s^2_{AdS_5}+\dd s^2_{\mathcal{M}^5}\Big),\qquad\qquad \dd s_{AdS_5}^2 =-\dd t^2 \cosh^2 r +\dd r^2 + \sinh^2 r\, \dd s_{S^3}^2\,,
\end{equation}
where $\dd s^2_{\mathcal{M}^5}$ is given by \eqref{xxx}, we 
 perform the rescaling
\begin{equation}\la{427}
x^-\to\frac{1}{L^2}\, x^-\,,\qquad r \to \frac{{1}}{L}r\,,\qquad \rho\to\frac{1}{L}\rho\,,\qquad a\to \frac{1}{L}a\,, \qquad \quad 
x^\pm \equiv \frac{1}{2}(t\pm \chi) \ . 
\end{equation}
At leading order in large $L$ this results in the following pp-wave metric
\ba \label{533}
\dd s_{10}^2=&- 4\dd x^+\dd x^- - \big(x^2+\rho^2\big)(\dd x^+)^2+ \dd x^i\dd x^i + V_0(\rho)^{-1}\,\dd \rho^2+\rho^2\Big[\sigma_x^2 + \sigma_y^2+ V_0(\rho)\, \sigma_z^2\Big] \,,\\
&\qquad \qquad V_0(\rho)=1-\tfrac{a^4}{\rho^4} \ , 
\ea 
where $x^i$ ($i=1,2,3,4$) are originating from the $AdS_5$ coordinates. 
This background is a pp-wave with the transverse space
being the product of $\mathbb{R}^4$ and the EH space with metric \rf{244}.
The $5$-form \rf{222} becomes 
\begin{equation}\la{566}
F_{5}= 4\, \dd x^+\wedge\big( \dd x^1\wedge \dd x^2\wedge \dd x^3\wedge \dd x^4 - \rho^3 \dd \rho \wedge \sigma_x\wedge \sigma_y\wedge \sigma_z\big)\,.
\end{equation}
One can check directly that this background solves the 10d supergravity equations 
 and preserves half of the maximal supersymmetry.

Alternatively, we may arrive at this background by first taking the Penrose limit 
of the orbifold 
$AdS_5\cross S^5/\mathbb{Z}_2$, resulting in an orbifolded pp-wave background 
\cite{Itzhaki:2002kh, Alishahiha:2002ev, Kim:2002fp, Floratos:2002uh, Sahraoui:2002sp}
with the transverse space being $\mathbb{R}^4\cross\mathbb{C}^2/\mathbb{Z}_2$. To resolve the orbifold singularity \cite{Floratos:2002uh}
 we may replace $\mathbb{C}^2/\mathbb{Z}_2$ by its EH resolution \rf{244}. One is then 
 to check that the resulting pp-wave metric satisfies the supergravity equations. As the EH metric is Ricci flat, this requires that the coefficient $H(x,\r)$ of the 
 $(\dd x^+)^2$-term in the pp-wave metric should satisfy the Laplace equation $\nabla^2 H=0$ 
 on the transverse space $\mathbb{R}^4\cross{\rm EH}_4$. Choosing 
 $H= x^2 + h(\rho)$ one finds that $h$ should solve the Poisson equation on the EH space
\begin{equation}
\nabla^2_{_{\text{EH}}} h(\rho)= 8 \ . 
\end{equation}
It has $h(\rho)=\rho^2$ as its simplest solution which reproduces \rf{533}.\footnote{Note that ref. \cite{Floratos:2002uh} discussed a different solution for $h(\rho)$.}

\def \nuk {\bar \nu}

Next, we also need to find a similar limit in the solution \rf{34},\rf{313} for the twisted sector mode.
The $e^{ik \chi}$ mode in \rf{313} should correspond to a particle moving fast along the $S^1$. We get 
\begin{equation}\la{431}
\Omega \ \to\ 
L^2\, e^{-2i \nuk x^-}\,
 \frac{e^{-\frac{\nuk}{2}\rho^2}}{\rho^2}\Big[\big(2+\nuk\rho^2\big)\sigma_x\wedge\sigma_y+i \nuk\rho^2 \dd x^+ 
 \wedge \sigma_z-\frac{2+\nuk\rho^2}{\rho}\dd\rho\wedge \sigma_z\Big]\,, 
 \qquad   \nuk \equiv {k\ov L^2} = {\nu \ov \a'} \ . 
\end{equation}
Indeed, starting with the analogue of the supergravity equations \rf{31},\rf{32} in the pp-wave background 
\rf{533},\rf{566} we get for the pp-wave analogue of the solution \rf{34},\rf{313} 
\begin{equation} \la{599}
A_2=\varphi(x^+,x^i)\,\Omega \,, \qquad \qquad 
\big(2i\nuk \partial_++\partial_i\partial^i -\nuk^2 x^2- 4 \nuk\big)\varphi=0\,.
\end{equation}
The $x^i$ dependence is found as in the harmonic oscillator problem. 
For the ground-state solution we get the dispersion relation corresponding to 
a particle with $m^2=\nuk^2-4 \nuk$. 

Like the original solution \eqref{313}, the pp-wave solution \eqref{431} diverges for $\rho\to0$ with the leading term being (cf. \rf{327}) 
\begin{equation}\la{5001}
\Omega \ \overset{\rho\to0}{\sim}\ L^2 e^{-
	2i \nuk x^-}\Theta_{\text{EH}}\,, 
\qquad \ \ \Theta_{\text{EH}}= \dd\big(\rho^{-2}\,\sigma_z\big)\ . 
\end{equation}
where $\Theta_{\text{EH}}$ 
is the anti-self-dual exact 2-form on the EH space with non-zero integral over the resolution cycle \cite{Eguchi:1978gw}. 
 Rescaling $A_2$ by $a^2$ we thus get 
\begin{equation}\la{5111}
A_2= L^2a^2\, \hat \varphi(x^-,x^+,x^i)\,\Theta_{\text{EH}}\,, \qquad \ \ \ \ \ \hat \varphi(x^-,x^+,x^i)= e^{-2i \nuk x^-} \, \varphi(x^+,x^i) \ , 
\end{equation} 
which satisfies the equations of motion up to terms of order $\rho^2$
\begin{equation}\la{436}
\nabla^2 A_{AB}+i F_{AB}\,^{CDE}\nabla_C A_{DE}=(\nabla^2 -2 i \partial_-)A_{AB}=\frac{\rho^2}{4}\partial_-^2A_{AB}\,.
\end{equation}
This is not an issue for the interpretation of $A_2$ as the origin of the twisted modes as according to the discussion 
in section \ref{sec 3.3} these modes localise near $\rho=0$. 
However, 
we will see below that the $\rho^2$-terms are still important to render the 
$\alpha'^3$-correction finite.

Let us return to the analysis of $\alpha'^3$-corrections \rf{420}
 in section \ref{sec 4.2}
now using the pp-wave background. 
The Weyl tensor corresponding to the pp-wave metric \eqref{533} (see appendix \ref{appendix D})
 splits into two parts 
\begin{equation}\la{5131}
\mathcal{C}=\mathcal{C}_{\text{EH}}+ \mathcal{C}_{\text{mix}}\,,\qquad\qquad \mathcal{C}_{\text{EH}}\sim \OO\big({\frac{a^4}{\rho^6}}\big)\,,\qquad \qquad \mathcal{C}_{\text{mix}}\sim \OO\big({\frac{a^4}{\rho^4}}\big)\,,
\end{equation} 
where $\mathcal{C}_{\text{EH}}=\mathcal{C}_{bcde}$ is the Weyl tensor of the EH space and $\mathcal{C}_{\text{mix}}$ has non-zero components of the form
\begin{equation}
(\mathcal{C}_{\text{mix}})_{+b+b}=\pm \frac{a^4}{\rho^4}\,,\la{5444}
\end{equation}
where $b=1,...,4$ is a Vierbein index of the EH space. 
The invariant in \rf{420} involves 
one power of the Weyl tensor and two powers of the $G_3=2A_2$ field,
which according to \rf{5111} is proportional to $\Theta_{\text{EH}} $. 
 One can check that contractions of 
 the form $\mathcal{C}_{\text{mix}}\Theta_{\text{EH}}\Theta_{\text{EH}} $ vanish. 
 We thus need to consider only contractions with $\mathcal{C}_{\text{EH}}=(\C_{bcde})$ with the relevant ones being 
\begin{equation}\la{439}
\C_{bcde}\, \Theta^{de}_{\text{EH}}=16\frac{a^4}{\rho^6}\Theta_{\text{EH},bc}\,,\qquad \qquad \C_{bdce}\,
 \Theta^{de}_{\text{EH}}=8\frac{a^4}{\rho^6}\Theta_{\text{EH},bc}\,.
\end{equation}
The EH Weyl tensor diverges as $\mathcal{C}_{\text{EH}}\sim\order{a^{-2}}$ in the orbifold limit $\rho\to a\to0$.
Therefore, an 
 insertion of $\mathcal{C}_{\text{EH}}$ should be accompanied by an additional factor
 proportional to $a^2$ or $\rho^2$ to get a finite correction. 
 According to \rf{436} 
 such a factor 
 may come from a $(\nabla^2-2i\partial_-) A_2$-term. 
\iffa 
 \footnote{Eq. \eqref{436} is an on-shell relation. However, we can always perform a field redefinition 
\begin{equation*}
A_2\to A_2 + \alpha'^3 \,\mathcal{O}\, [\text{E.o.M.}] A_2
\end{equation*}
proportional to the equations of motion. This allows us to explicitly cancel off-shell terms in the $\alpha'$-corrections. One might argue that this invalidates our entire discussion, as we are trying to determine the normalisation of the action, which is itself proportional to the equations of motion. However, here we are saved by the observation \eqref{436} that in our approximation the equations of motion are in fact not satisfied completely and the action remains finite on-shell.}
\fi

We can now specify more explicitly the conjectured 
 structure of the $\a'^3$-invariant in \rf{420} required to 
reproduce the leading $k^3$-term in \rf{411}. 
Staring with the term 
$F^{ABCDE}\mathcal{C}_{DE}\,^{FG} \nabla_A \bar G_{BC}\,^H\square G_{FGH}$ 
we need to add to it other terms with a smaller number of 
covariant derivatives in order to get the combination 
$ \nabla^2-2i\partial_-$ and
 to reproduce the right spectrum.
 This is achieved by starting with 
\begin{equation}\la{441}\begin{split}
\mathcal{L}_8\ \sim \ 
&F^{ABCDE}\mathcal{C}_{DE}\,^{FG}\ \Big( i\nabla_A \bar G_{BC}\,^H \nabla^2 G_{FGH} 
+ \ha 
F_{FG}\,^{HIJ} \nabla_A\bar G_{BC}\,^K \nabla_H G_{IJK}\\
&+ F_{FG}\,^{HIJ} \bar G_{ABC}\nabla^2 G_{HIJ}
-\tfrac{i}{2}F_{FG}\,^{HIJ} F_{IJ}\,^{KMN}\bar G_{ABC} \nabla_H G_{KMN}+ \text{c.c.}\Big)\,.
\end{split}
\end{equation}
Using the equations of motion and integrating over the EH space 
we then find the following $\a'^3$-correction to the 6d effective action (cf. \rf{4122}) 
\footnote{Note that any contraction of $F_5$ with $\Theta_{\text{EH}}$, $\mathcal{C}$ and another $F_5$ can only involve indices from the EH space, leaving at least the $+$ index of $F_5$ uncontracted. 
This then requires a contraction with $\nabla_-$.}
\begin{equation}\la{444}
 \Delta S_6\ \sim \ \alpha'^3\zeta(3)\, \Vv \, 
\int_{AdS_5\cross S^1}\dd^5 x\, {\dd\chi} 
\,\sqrt{-g_6}\, \Big( i \nabla^A\bar{\hat \varphi} \nabla_-^3 \nabla_A\hat \varphi + 
 4 \bar{\hat \varphi} \nabla_-^4 \hat \varphi+ \text{c.c.}\Big)\,.
\end{equation}
This correction has the same structure as the leading-order action in \rf{18} (with $\hat \varphi= \beta + i \gamma$ as in \rf{323}) but with 
 an extra insertion of the operator $\nabla_-^3$. It thus corresponds to the $\del_\chi^3$-term in \rf{424}. 
 Acting on $\hat \vp$ in \rf{5111} it produces the expected $\nu^3$-term in the 2-point function \rf{411}.

\iffa 
This is just the 6d action \eqref{18} with an insertion of the operator $\tfrac{1}{2}(i\nabla_-)^3$, which evaluates to $\frac{1}{2}(2\nu)^3$ on our solution and thus reproduces the operator $c(\partial_\chi)$ \eqref{425} we expected to find. To make contact with \eqref{6d}, we need to adjust the integration volume $\mathcal{V}$ in the orbifold limit and reinstate the factor $\mathcal{V}_0$ from \eqref{413}. Any additional constant factors arising during this procedure may have to be anticipated by adjusting the normalisation of \eqref{441}.
\fi

Let us note that the suggested structure \eqref{441} of the required $\a'^3$-term is not unique. 
 For example we could use other contractions of internal indices, like the second option in \eqref{439}, 
 or a different ordering of the $F_5$ and $\mathcal{C}$ factors. The structure of ${\mathcal L}_8$ in \rf{415} 
 dictated by supersymmetry should be a combination of all such terms that leads to \rf{444} with the right overall coefficient to match the one in \rf{411}.

\iffa 
It should be noted, however, that we expect no further corrections to the 6d normalisation at this order in $\alpha'$. This suggests that for example terms involving $\mathcal{C}^2$ should either vanish or contribute similar terms to \eqref{441}. In fact, given that $\mathcal{C}_{\text{EH}}^2$ diverges as $\tfrac{a^8}{\rho^{12}}$ and accounts for 4 derivatives, we have no way of cancelling this divergence with $\rho^2$-contributions from \eqref{436}. We therefore conjecture that an algebraic cancellation occurs for these terms. $\mathcal{C}_{\text{EH}}\mathcal{C}_{\text{mix}}$ is linear in $\mathcal{C}_{\text{mix}}$ and therefore cannot contract with $\Theta_{\text{EH}}^2$, which leaves us with $\mathcal{C}_{\text{mix}}^2$. This tensor has to be contracted with four $\nabla_-$-factors and thus contributes the same 6d correction as the flux contribution in the last line of \eqref{441}. 
\fi 

\iffa 
$\mathcal{C}^3$-terms are even more constrained, such that no terms involving $\mathcal{C}_{\text{mix}}$ are possible. However, we again require some algebraic cancellation to get rid of the divergence from $\mathcal{C}_{\text{EH}}^3$. An indication for such cancellations arises from the fact that the known $\mathcal{R}^4\sim\mathcal{C}^4$ term in \eqref{415} vanishes on the background \eqref{533}, suggesting that the index structures contracting multiple Weyl tensors are indeed conductive to cancellations.
\fi

\subsection{Back to $AdS_5\cross \mathcal{M}^5$}\label{sec5.2}

Let us now see how to reproduce the $k^3$-term in \rf{411} without first taking the pp-wave limit, i.e.
 by starting with 
the string effective action expanded near $AdS_5\cross \mathcal{M}^5$ 

\iffa 
We now turn to the $AdS_5\cross \mathcal{M}^5$ model before taking the large-$k$ limit and see whether we can treat it analogously. It turns out that upon expansion of the relevant $AdS_5\cross \mathcal{M}^5$ quantities in small $\rho$, the leading order terms match the large-$k$ quantities. Given that only the leading order terms are relevant for the determination of $\alpha'$-corrections, we immediately reproduce the $k^3$ term. Subleading corrections in $k$ now arise because subleading terms in the $\rho$-expansion of $AdS_5\cross \mathcal{M}^5$ quantities break the cancellation structure of the large-$k$ limit.
\fi

The Weyl tensor $AdS_5\cross \M^5$ written in Vielbein components 
(see appendix \ref{appendix D}) can be split, like in \rf{5131}, into two parts 
\begin{equation}\la{519}
\mathcal{C}=\mathcal{C}_{\text{int}}+ \mathcal{C}_{\text{mix}}\,,\qquad \mathcal{C}_{\text{int}}\sim \order{\frac{\sin^4 a}{\sin^6 \rho}}\,,\qquad \mathcal{C}_{\text{mix}}\sim \order{\frac{\sin^4 a}{\sin^4 \rho}}\,,
\end{equation} 
where $\mathcal{C}_{\text{int}}$ resembles the EH Weyl tensor $\mathcal{C}_{\text{EH}}$ for small $\rho$ 
\begin{equation}
\mathcal{C}_{\text{int}}\overset{\rho\to0}{\sim}\mathcal{C}_{\text{EH}}+\order{\frac{a^4}{\rho^4}}\ , \la{5200}
\end{equation}
and $\mathcal{C}_{\text{mix}}$ only has entries of the form
\begin{equation}\la{521}
(\mathcal{C}_{\text{mix}})_{\chi b \chi b}=\pm\frac{\sin^4 a}{\sin^4\rho}\,.
\end{equation}
Expanding the approximate solution for $\Omega= e^{ik\chi} \Theta $ in \rf{327} for small $\rho$ 
yields $\Theta_{\text{EH}}$ as the leading $\OO(\rho^{-2})$-term in $\Theta$
\begin{equation}
\Omega= e^{ik\chi} \Theta \ 
\overset{\rho\to0}{\sim}\ e^{ik\chi}\, \Theta_{\text{EH}}+\order{\rho^0}\,. \la{5223}
\end{equation}
 We check that like in the pp-wave limit in \rf{436}, the field 
 $A_2=\hat \varphi(x,\chi)\Theta$ in \rf{327a} satisfies the equations of motion \rf{32} 
 up to $\order{\rho^2}$-terms
\begin{equation}\la{449}
\nabla^2 A_{AB}+i F_{AB}\,^{CDE}\nabla_C A_{DE}=-k^2\rho^2 A_{AB}+\order{\rho^4 A}= \rho^2\partial_\chi^2 A_{AB}+\order{\rho^4 A}\,.
\end{equation}
Given that only the leading term in $\rho\to0 $ should contribute to the relevant part of the action,
 we may use the same $\alpha'^3$-combination as in \eqref{441}, now starting with 
 the $AdS_5\cross \mathcal{M}^5$ background. 
It evaluates to the same 6d correction \eqref{444} 
 reproducing again the $k^3$-term in \eqref{411}. 

 In addition, we expect also other corrections that should correspond to $\zeta(3)$-terms with lower
 powers of $k$ in \rf{411}. Various cancellations that prevent divergence of the $\alpha'^3$-corrections in the 
 pp-wave limit should still occur here, but only up to subleading finite terms.\foot{Consider, for example, the $\mathcal{C}_{\text{int}}^2$-term which diverges as $\tfrac{a^8}{\rho^{12}}$. 
 We expect this divergence to cancel algebraically up to order $\tfrac{a^8}{\rho^{10}}$, where finite contributions can arise when multiplied with the operator in \eqref{449}. These finite contributions did not arise in the large-$k$ limit, so they 
 should be due to the $\rho^2$-corrections to the $AdS_5\cross \mathcal{M}^5$ quantities.} 
 To illustrate the cancellation structure we expect, let us consider the contraction
\begin{equation}\la{5244}
\sqrt{-g}\,\mathcal{C}^{ABCD}\mathcal{C}_{CD}\,^{EF}(\bar\Omega_{AB}\Omega_{EF}-8\bar\Omega_{AE}\Omega_{BF})\,.
\end{equation}
As this term involves two Weyl tensors $\mathcal{C}$ and two 2-forms $\Omega$,
 we would expect a small-$\rho$ divergence of order $\tfrac{a^{12}}{\rho^{17}}$.
 In fact, this contraction vanishes on the pp-wave background of section
 \ref{sec5.1} and behaves as $\tfrac{a^{12}}{\rho^{13}}$ in the $AdS_5\cross \mathcal{M}^5$ case. 
 Supplementing \rf{5244} with other terms involving factors of derivatives and $F_5 $
 we may be able to build an invariant like \eqref{441} that
 reproduces the $k^2$-term in \eqref{411} but vanishes when evaluated on the pp-wave background. 

At this point, the appearance of such cancellations in an eventual complete description of the $\alpha'^3$-terms is speculative. However, let us mention that precisely this 
cancellation pattern is observed in  the 
explicitly known 
$\alpha'^3\zeta(3) \mathcal{R}^4$ term in \eqref{415}:
 despite involving  four powers of the Weyl tensor it 
 vanishes on the pp-wave background and is finite in the $\rho\to a\to 0$ limit on $AdS_5\cross \mathcal{M}^5$. 

\section{Concluding remarks}\label{sec6}

In this paper we suggested a strategy 
of matching the large-coupling expansion of twisted sector correlators in planar 
 4d $\mathcal{N}=2$ superconformal 
$SU(N)\cross SU(N)$ quiver gauge theory
 to $\alpha'$-corrections in the dual orbifold string theory. Specifically, we considered 
 the 2-point function of the 
 twisted sector operators $T_k$ in \eqref{ops}. 
 
 The corresponding twisted sector string modes 
 localise on the fixed 
 $AdS_5\cross S^1$ subspace. 
To access these localised modes, we proposed an explicit resolution of the $S^5/{\mathbb Z}_2$ orbifold singularity
 \eqref{xxx}, represented 
 by a (non-supersymmetric)
 solution of 10d type IIB supergravity. 
  This solution is not  fully regular  but introduces 
 a non-trivial 2-cycle on which the 2-form supergravity fields can ``wrap''. 
 
 These extra modes should effectively represent the lightest 
 twisted sector string modes that should be present in the first-principles string theory approach
 (which is possible in flat space but is not directly available 
 in the $AdS_5 \cross S^5/{\mathbb Z}_2$ case). 
 We derived an effective 6d action 
 for these 2-form modes following from the supergravity action expanded near the resolved background. 
 An analogous treatment for other light modes in the twisted sector should be possible  too (cf. \cite{Gukov:1998kk}). 

We then suggested how the inclusion of the $\alpha'^3$-corrections to the type IIB effective action may allow one to match 
the subleading term in the localisation result \rf{3}. 
Our discussion remained 
at a qualitative level 
 due to lack of knowledge about the full expression for the supersymmetric completion of the 
 $\a'^3 R^4$ invariant. We pointed out the important simplification that happens in the 
 large $R$-charge limit $k\to\infty$ when the resolved orbifold background simplifies to a pp-wave one. 

It would be interesting to extend our approach to the 3-point function \rf{4} and understand the string theory origin of the 
relation between the 2-point and 3-point coefficients mentioned in footnote \ref{f4}. 
Another possible extension is to the case of the $AdS_5 \cross S^5/{\mathbb Z}_{\rm L}$ 
orbifold dual to the $\rm L $-node quiver in which case the generalisation of the expansions in \rf{3},\rf{4} 
where recently found (see \ci{Beccaria:2023qnu} 
and refs. there). 
Ref. \cite{Beccaria:2023qnu} observed interesting simplifications
 occurring 
  in the large-$\rm L$ limit correlated with taking $k$ or $\l$ large; 
  this may be suggesting 
 the existence of a similar well-defined 
 limit on the string theory side. 
 
 \iffa 
A somewhat orthogonal direction is to study the deformation that changes the relative gauge couplings of the two $SU(N)$-nodes in the gauge theory quiver. This deformation is supposed to be generated by some finite $B_2$-potential wrapping the resolution cycle. As such, it has no immediate effect on our current discussion, as a finite shift of the $B_2$-field has no effect on the equations of motion. In fact \cite{Klebanov:1999rd} suggests that some finite $B_2$-field is already turned on at the orbifold point. It would be interesting to develop some effective supergravity description of the deformed model and to identify observables that are affected by this shift.
\fi

\section*{Acknowledgements}
We would like to thank Matteo Beccaria, Nikolay Bobev, Jerome Gauntlett, Gregory Korchemsky, Elli Pomoni 
 and Bogdan Stefanski for helpful and encouraging discussions.
 TS acknowledges funding by
the President's PhD Scholarship of Imperial College London.
AAT is supported in part by the STFC Consolidated Grants ST/T000791/1 and ST/X000575/1.

\newpage

\appendix 
\section{String spectrum for 
a 
flat-space orbifold}\label{appendix A}

Here we review some facts about the superstring spectrum on
a 
 particular    $R^{1,6}\cross \mathbb{C}^2/\mathbb{Z}_2$ orbifold
 (see \ci{Dixon:1985jw,Becker:2006dvp,Douglas:1996sw}).
Starting with $\mathbb{R}^{1,9}$ we identify coordinates 
as $(x_5,x_6,x_7,x_8)\sim(-x_5,-x_6,-x_7,-x_8)$.\footnote{We can also describe this as simultaneously rotating two $\mathbb{C}$-planes by an angle $\pi$.} 
We shall use the Green-Schwarz formulation in light-cone gauge 
 with the action ($i=1,2...,8$) 
\begin{equation}
\mathcal{S}=\frac{1}{\pi\alpha'}\int\dd^2\sigma~ \Big(\partial_+x^i\partial_-x^i + iS_R^\alpha\partial_+S_R^\alpha+iS_L^\alpha\partial_-S_L^\alpha\Big)\ .
\end{equation} 
We are to 
 keep only states that are invariant with respect to the $\mathbb{Z}_2$-symmetry. 
 In particular, states localised at the singularity (i.e. with $x_0^a=p_0^a=0$, where $a\in\{5,6,7,8\}$)
 can only have even numbers of $\mathbb{Z}_2$-odd excitations, as, for example, the bosonic modes $\alpha_{-n}^a$ and $\tilde\alpha_{-n}^a$. To analyse the fermionic modes, we need to decompose the two $SO(8)$-spinors $S_L$ and $S_R$, which in type IIB theory have equal chirality, according to the splitting rules
\ba \label{decomp}
&\qquad SO(8)\to SU(2)\cross SU(2)\cross SU(2)\cross SU(2)\,, \\
\mathbf{8_v}\to (\mathbf{2,2,1,1})\oplus&(\mathbf{1,1,2,2})\,,\quad 
\mathbf{8_s}\to (\mathbf{2,1,2,1})\oplus(\mathbf{1,2,1,2})\,,\quad 
\mathbf{8_c}\to (\mathbf{1,2,2,1})\oplus(\mathbf{2,1,1,2})\,.\no 
\ea
The orbifold action in this decomposition is represented by 
\begin{equation}
\Gamma= \mathbb{1}_2\otimes \mathbb{1}_2\otimes \mathbb{1}_2\otimes \begin{pmatrix}
-1 & 0\\
0 & -1
\end{pmatrix}\,.
\end{equation}
We interpret the first two quantum numbers as determining the representation in the 4d-space spanned by $x^i$ with $i\in{1,2,3,4}$. The remaining untwisted $SU(2)$ will become an $\mathcal{R}$-symmetry in an eventual compactification.

The localised states belong to two distinct sectors, the untwisted one, which is just the $\mathbb{Z}_2$-invariant part of the usual type IIB spectrum, and the twisted one which closes up to a $\Gamma$-transformation. Let us analyse them in turn.

 \textbf{ Untwisted sector:}
	
\noindent	We can follow the usual IIB construction and use \eqref{decomp} to split
	\begin{equation}
	(\mathbf{8_c}\oplus\mathbf{8_v})\to \Big((\mathbf{2,2,1,1})\oplus(\mathbf{1,2,2,1})\Big)_1\oplus\Big((\mathbf{1,1,2,2})\oplus(\mathbf{2,1,1,2})\Big)_{-1}\,,
	\end{equation}
	where the subscript denotes the eigenvalue under $\Gamma$. When we combine left- and right-movers, we only keep $\Gamma$-invariant states, so we can ignore cross-terms and drop the $\mathbf{3}$-representations of the last $SU(2)$. This results in the spectrum 
	\begin{equation}\begin{split}
	I:& \quad (\mathbf{3,3;1})\oplus (\mathbf{2,3;2})\oplus(\mathbf{2,3;2})\oplus(\mathbf{1,3,1})\oplus(\mathbf{1,3;1})\oplus(\mathbf{1,3;3})\,,\\
	II:& \quad (\mathbf{3,1;1})\oplus (\mathbf{2,1;2})\oplus(\mathbf{2,1;2})\oplus(\mathbf{1,1;1})\oplus(\mathbf{1,1;1})\oplus(\mathbf{1,1;3})\,,\\
	III:& \quad (\mathbf{3,1;1})\oplus (\mathbf{2,1;2})\oplus(\mathbf{2,1;2})\oplus(\mathbf{1,1;1})\oplus(\mathbf{1,1;1})\oplus(\mathbf{1,1;3})\,.
	\end{split}
	\end{equation}
	Here, we dropped the last $SU(2)$ factor (all fields are in the representation $\mathbf{1}$) and separated the $\mathcal{R}$-symmetry. The three sets form representations of 6d $\mathcal{N}=(2,0)$ supergravity, namely the gravity multiplet $I$ and two tensor multiplets $II$ and $III$. Alternatively, we could have arrived at these multiplets 
	 by expanding the known type IIB spectrum in $SU(2)$ representations and projecting upon the $\mathbf{1}$ of the fourth $SU(2)$ factor.
	
 \textbf{Twisted sector}:
	
\noindent	In the twisted sector, half of the bosonic and fermionic oscillators have to be glued with 
antisymmetric boundary conditions. In general, such twisted boundary conditions can affect the normal ordering constant, but here it receives equal contributions from fermions and bosons and therefore vanishes. However, only the fermions with symmetric boundary conditions have zero modes, so at the massless level, the usual ground-ground-statestate degeneracy is 
generated only by one spinor $\hat S_0^\alpha$ in the representation $(\mathbf{2,1,2,1})$. Starting with a $\Gamma$-charged state $\ket{a}\in (\mathbf{1,1,2,2})$, we can act with $\hat S_0^\alpha$ and create 
	\begin{equation}
	\dot{\ket{\beta}}=\Gamma^a_{\dot{\beta}\alpha}\hat{S}_0^\alpha\ket{a}\in (\mathbf{2,1,1,2})\,.
	\end{equation}
	Another action with $\hat S_0^\alpha$ leads back to $(\mathbf{1,1,2,2})$. Therefore the massless states generated in both left- and right-moving sector furnish the reduced representation
	\begin{equation}
	(\mathbf{1,1,2,2})\oplus(\mathbf{2,1,1,2})\,.
	\end{equation}
	Combining the left- and right-moving states we get 
	\begin{equation}
	T: \quad (\mathbf{3,1;1})\oplus (\mathbf{2,1;2})\oplus(\mathbf{2,1;2})\oplus(\mathbf{1,1;1})\oplus(\mathbf{1,1;1})\oplus(\mathbf{1,1;3})\,,
	\end{equation}
	which is another tensor multiplet of 6d $\mathcal{N}=(2,0)$ supergravity.

\section{Comments on $\alpha'^3$-terms in the type IIB string effective action }\label{appendix B}

The familiar $R^4 $-term in the tree level type II superstring effective Lagrangian may be written as\foot{ 
We shall ignore the dependence on the dilaton and set $\epsilon_8\epsilon_8 =-\frac{1}{2}\epsilon_{10}\epsilon_{10}$.}
\begin{equation}\label{R4}
\mathcal{L} = \frac{\alpha'^3}{3\cdot 2^{11}} \, \zeta(3)\, \Big(t_8t_8-\frac{1}{4}\epsilon_8\epsilon_8\Big) \,\mathcal{R}^4\, . 
\end{equation}
Here the $t_8t_8$-term is fixed from 4-graviton amplitude \ci{Gross:1986iv} while the presence of the $\epsilon$-term 
 may be deduced from sigma model considerations
 \cite{Grisaru:1986vi,Freeman:1986zh} or by computing a 5-point amplitude (see a review in \cite{Liu:2013dna}).
 Other $\a'^3$-terms should be related to \rf{R4} by supersymmetry. We are interested, in particular, 
 in the terms quadratic in $G_3= H_3 + i F_3$ but they appear to be not known completely. 
 
 Ref. \cite{Liu:2019ses} 
 discussed some subset of terms with $ \mathcal{R}^4 \to \mathcal{R}^4 + 6 \mathcal{R}^2 \abs{\nabla G}^2 + ...$
 in \rf{R4} and extra $\abs{\nabla G}^4$ that have different index structure. 
 Another approach followed in \cite{Liu:2022bfg} was to use contraction tensors $t_{m}$ which are expected to arise 
 from 10d superspace integrals. In particular, they considered the following subset of terms ($c_n$ and $c_5$ are numerical constants)
 \ba\label{b15}
&\mathcal{L}\ \sim \ \sum_{n=0}^4 c_n \, t_{24}\, G_3^{2n}\bar{\mathcal{R}}_{(6)}^{4-n}\,,\\
\label{R_6}
\bar{\mathcal{R}}_{ABCDEK}=&\tfrac{1}{8}g_{CK} \mathcal{C}_{ABDE}+ \tfrac{i}{48}\nabla_A F_{BCDEK} \\
& +\tfrac{1}{768}\Big(F_{ABCLM}F_{DEK}\,^{LM}-3F_{ABKLM}F_{CDE}\,^{LM}\Big)+ c_5G_{ABC}\bar G_{DEK}\,,
\ea 
where $\mathcal{C}_{ABCD}$ is the Weyl tensor and $F_5$ is the RR 5-form. 
We refer to \cite{Liu:2022bfg} for details. 

\iffa 
Another benefit of the higher index structures is a simplification of the quintic terms found in \cite{Liu:2019ses}. The quartic and quintic terms involving $R^3$ and $G_3^2$ are summarised as 
\begin{equation}\begin{split}
e^{-1}\mathcal{L}\supset \frac{\alpha'^3}{3\cdot 2^{12}}\bigg[&f_0(\tau,\bar{\tau})t_{16}\bar{\mathcal{R}}^4+\frac{3}{2}\Big(f_1(\tau,\bar{\tau})t_{18}G_3^2R^3+f_{-1}(\tau,\bar{\tau})t_{18}\bar G_3^2R^3\Big)\\
&+f_0(\tau,\bar{\tau})\Big(-\frac{1}{3}\epsilon_9\epsilon_9+2\tilde t_8\tilde t_8-\frac{1}{2}\epsilon_8\epsilon_8-t_{18}\Big)\abs{G_3}^2R^3 \bigg]\,,
\end{split}
\end{equation}
where the $\tilde t_8 \tilde t_8 $ contraction is to be understood as $t_8^{ij\dots}t_8^{kl\dots}G_{ik}\,^a\bar{G}_{jak}\dots$\,, and similar for $\epsilon_8$. Most noticeably, the first term with $t_{16}$ structure captures \eqref{R4} and would have to be expanded to the full \eqref{4pt} to capture all quartic terms involving $G_3$.
\fi

\section{Solution of the 2-form equations of motion}\label{appendix C}
In this appendix we solve  the equation 
\eqref{onset} by imposing  the spherically symmetric Ansatz \eqref{311}.
 At finite value of the resolution parameter 
  $a$ it is useful to define $\td k =p^{-1} {k} $ such that $k\chi= \td k \td \chi$ (cf. \eqref{modforms},\eqref{216}). In the orbifold limit $a\to 0$ the distinction
  between $\td k$ and $k$  disappears.

Demanding closure of the 3-form $\omega$ results in the equation
\begin{equation}\label{closure}
2f_1(\rho)-i\td kf_2(\rho)+f_3'(\rho)=0\,.
\end{equation}
Eq.\eqref{onset} takes the form of 3 coupled ordinary differential equations
\begin{align}
-i M f_1(\rho)&= \partial_\rho\big[\cot\rho\, V(\rho)\,f_2(\rho)\big]+ i \td k \frac{1}{\cos\rho\sin\rho} \, f_3(\rho)\,,\\
-i M f_2(\rho)&= \partial_\rho\big[\tan\rho\,f_1(\rho)\big]+ \frac{2}{\cos\rho\sin\rho}\,f_3(\rho)\,,\\
-i M f_3(\rho)& = i \td k \tan\rho \,f_1(\rho) - 2 \cot\rho\, V(\rho)\, f_2(\rho) \,,
\end{align}
which  are consistent  with \eqref{closure}. We can solve algebraically for $f_1$ and $f_2$ 
\ba
& f_1(\rho)=-\frac{\td kM \tan\rho \,	f_3(\rho)+
2 V(\rho) f_3'(\rho)}{W(\rho)}\,,\qquad \  \ \ 
f_2(\rho)=i\frac{2M\tan\rho\, f_3(\rho)-\td k \tan^2\rho\, f_3'(\rho)}{W(\rho)}\,,
\\  \label{W}
&\qquad \qquad \qquad\qquad  W(\rho)=\td k^2\tan^2\rho+4 V(\rho)\,.
\ea
We then  get  a second order linear ODE for $f_3$ of the form
\ba\label{ODE}
& f''_3(\rho)+P(\rho)f'_3(\rho)+Q(\rho)f_3(\rho)=0\,,
\\
&P(\rho)=\frac{1}{\cos{\rho}\sin{\rho}}-\frac{W'(\rho)}{W(\rho)}+ \frac{V'(\rho)}{V(\rho)}\,,\\
&Q(\rho)=\frac{1}{V(\rho)}\Big[-\frac{W(\rho)}{\sin^2\rho\,}+M^2-\td k M\Big(\frac{\tan\rho}{2}\frac{W'(\rho)}{W(\rho)}-\frac{1}{\cos^2\rho}\Big)\Big]\,.\la{c9}
\ea
Introducing $\gf(\rho)$  defined as 
\begin{equation}\la{c11}
f_3(\rho)={\rm exp} \Big[-\frac{1}{2}\int P(\rho)\Big]\, \gf(\rho)= \sqrt{\cot\rho\frac{ \,W(\rho)}{\, V(\rho)}}\ \gf(\rho)\,,
\end{equation}
we get  for it  the Schr\"odinger type equation \eqref{312}, with the potential
\begin{equation}\label{potential}
U(\rho)=\frac{1}{4}P^2(\rho)+\frac{1}{2}P'(\rho)-Q(\rho)\,.
\end{equation}
$U(\r)$  has  the form of a smooth well between two poles at $\rho=0$ and $\rho=\frac{\pi}{2}$  ($\epsilon \to 0$)
\begin{equation}
U(0+\epsilon)\sim\frac{15}{4 \epsilon^2}+\order{1}\,,\qquad U\Big(\frac{\pi}{2}-\epsilon\Big)\sim \frac{k^2-\frac{1}{4}}{\epsilon^2}+\order{1}\,.
\end{equation} 
For $a>0$ we also find a pole at $\rho=a$ as is evident  from 
  \eqref{314} and is  illustrated in \ref{figure2}.

Imposing appropriate boundary conditions restricts the value of the 
 parameter  $M$  in \rf{c9}
to a  discrete set  for every given $\td k$. 

For  $a=0$ the spectrum of normalisable modes corresponds to a subsector of the usual Kaluza-Klein spectrum on the (orbifolded) sphere \cite{Kim:1985ez}.
 It is instructive to study this case in more detail.

\subsection*{Solutions for $a=0$}

At $a=0$ the coefficient functions $P$ and $Q$ take the  form
\begin{equation}
P_0(\rho)=\frac{1}{\cos\rho\sin\rho}\Big(1-2k^2\frac{\tan^2\rho}{W_0(\rho)}\Big),\qquad Q_0(\rho)=-\frac{W_0(\rho)}{\sin^2\rho}+M^2+\frac{4kM}{W_0(\rho)\cos^2\rho}\,,
\end{equation}
where $W_0(\rho)=k^2\tan^2\rho+4$ is the value of  $W(\rho)$ in 
\eqref{W} at $a=0$. We can put \eqref{ODE} in the Sturm-Liouville form
\begin{equation}\label{SL}
\partial_\rho\Big[\frac{\tan\rho}{W_0(\rho)}\partial_\rho f_3(\rho)\Big]+\frac{\tan\rho}{W_0(\rho)}Q_0(\rho)f_3(\rho)=0\,.
\end{equation}
The  corresponding norm  of $f_3$ is then 
\begin{equation}
\norm{f_3}^2=\int_0^{\frac{\pi}{2}}\dd \rho\,\frac{\tan\rho}{W_0(\rho)} f_3(\rho)\bar{f}_3(\rho)\,.
\end{equation}
Normalisability is guaranteed if $f_3$ vanishes at both singularities. 
Close to the singularities, \eqref{ODE} takes the form of the Bessel equation, and we can extract the leading asymptotics  of the solution as 
\begin{equation}
f_3(\rho)\overset{\rho\to0}{\sim}c \sin^2\rho+ \tilde c \sin^{-2}\rho\,,\qquad\qquad 
f_3(\rho)\overset{\rho\to\frac{\pi}{2}}{\sim} d\cos^{k}\rho+ \tilde d \cos^{-k}\rho\,.\\
\end{equation}
Then  normalisability restricts $\tilde{c}=\tilde{d}=0$. 
Introducing $h(\rho)$ defined by 
\begin{equation}
f_3(\rho)=\sin^2\rho\cos^{k}\rho \, h(\rho)\,,
\end{equation}
we get for it  the following equation
\begin{equation}
\partial_\rho\Big[\mu(\rho)\partial_\rho h(\rho)\Big]+\mu(\rho)(M-k-2)\Big[(M+k+2)+\frac{4k}{W_0\cos^2\rho}\Big]\, h(\rho)=0\,, \la{c18}
\end{equation}
where 
\begin{equation}
\mu(\rho)=\frac{\cos^{2k-1}\rho\sin^5\rho}{W_0(\rho)}\,,
\end{equation}
is again the relevant measure that defines  the Sturm-Liouville norm. 
We shall assume Neumann boundary conditions.

The potential in \rf{c18} vanishes for $M=k+2$, 
in which case this equation is  solved by constant $h(\rho)$.
 This  corresponds to the effective 5d masses 
\begin{equation}\label{KK-spectrum}
m^2=(k+2)(k+6)\,,\qquad \bar{m}^2=(k+2)(k-2)\,,
\end{equation}
which match the spectrum found in \cite{Kim:1985ez}. 
We also  find normalisable modes at every successive value $M=k+2(n+1)$, where $n\in\mathbb{N}_0$ denotes the number of zeros of the function $h(\rho)$. This is the KK tower of 2-form excitations obeying our symmetry requirements.
 If we try to extend this tower to $n<0$ we instead find that the boundary asymptotics change to $f_3(\rho)\sim \sin^{-2}\rho$. We also find a tower of non-normalisable solutions at $M=k-2n$, $n\in\mathbb{N}_0$.

\subsection*{Twisted sector solutions}

Turning to the discussion of  solutions  corresponding  to the 
twisted sector, let us now pretend that the boundary conditions at $\rho=0$ can be violated and $f_3\sim \sin^{-2}\rho$ is still a valid solution. We can then again 
extract a factor  from  $f_3$  by  setting 
\begin{equation}
f_3(\rho)=\frac{\cos^{k}\rho }{\sin^{2}\rho}\, \tilde h(\rho)\,,
\end{equation}
which results in the following  equation for $\td h$
\ba\label{twistedOPE}
\partial_\rho\Big(\tilde\mu(\rho)\partial_\rho \tilde h(\rho)\Big) &+\tilde\mu(\rho)(M+k-2)\Big((M-k+2)+\frac{4k}{W_0\cos^2\rho}\Big)\tilde h(\rho)=0\,,\\
&
\tilde\mu(\rho)=\frac{\cos^{2k-1}\rho\sin^{-3}\rho}{W_0(\rho)}\,,
\ea
which is again solved with Neumann boundary conditions.

To make contact with 
the relevant solution  in \eqref{327}, we need to require that the component
\begin{equation}\label{chizpart}
\Omega_{\chi z}=V(\rho)\, \cot\rho\,  f_2(\rho)\ \overset{a=0}{=}\ \cot\rho\,  f_2(\rho)
\end{equation}
is finite at  $\rho\to 0$. This singles out the non-normalisable solution at $M=-k$ given by\footnote{From the previous perspective, this is a diagonal cross-section through the tower $M=k-2n$ at $n=k$.}
\begin{equation}\label{twisted solution}
f_3=\frac{\cos^{k}\rho}{\sin^{2}\rho}\, \big(2 \cos^2\rho+ k\sin^2\rho\big)\,,
\end{equation}
which reproduces the spectrum of twisted states dual to the $T_k$  operators in \eqref{ops}.
 The other two functions in the full Ansatz \eqref{311} are given by
\begin{align}
f_1(\rho)=\frac{\cos^{k+1}\rho}{\sin^3\rho}\big(2+k \sin^2\rho\big)\,,\qquad \qquad
f_2(\rho)=i k\cos^{k-1}\rho\sin\rho\,.
\end{align}
Then
\begin{equation}
\Omega=e^{ik\chi}\Big(\tan\rho\, f_1(\rho)\, \sigma_x\wedge\sigma_y +\cot\rho\, f_2(\rho)\, \dd\chi\wedge\sigma_z -\frac{f_3(\rho)}{\cos\rho\sin\rho} \, \dd\rho\wedge\sigma_z\Big)\,,
\end{equation}
takes the form  given in \rf{313}.

\section{Weyl tensors in Vielbein basis}\label{appendix D}

\subsection*{pp-wave background}
We can write  the metric \eqref{533} in  a partial Vielbein basis \ba
&e^{i}=\dd y^i\,,\qquad e^5=\frac{1}{\sqrt{V_0(\rho)}} \dd \rho\,,\qquad e^6= \rho\, \sigma_x\,,\qquad e^7= \rho \,\sigma_y\,,\qquad e^8 = \rho \sqrt{V_0(\rho)}\,\sigma_z\,,\\
&\dd s_{10}^2=- 4\dd x^+\dd x^- - \Big(y^2+\rho^2\Big)(\dd x^+)^2+\sum_{A=1}^8\dd e^A\dd e^A \,. \ea
The corresponding components of $F_5$ in \rf{566} are then 
$F_{+1234}=F_{+5678}=4$.
In this basis, the  non-vanishing   components  of the Weyl tensor $\mathcal{C}$ are
\ba
&\mathcal{C}_{5656}=\mathcal{C}_{5757}=\mathcal{C}_{6868}
=\mathcal{C}_{7878}=-\mathcal{C}_{5678}=\mathcal{C}_{5768}=-2\frac{a^4}{\rho^6}\,, \label{Weyl1}\quad 
\mathcal{C}_{5858}=\mathcal{C}_{6767}=-\mathcal{C}_{5867}=4\frac{a^4}{\rho^6}\,, \\
&\mathcal{C}_{+5+5}=-\mathcal{C}_{+6+6}=-\mathcal{C}_{+7+7}=\mathcal{C}_{+8+8}=\frac{a^4}{\rho^4}\,,\label{Weyl2}
\ea
and other components
 related by the usual symmetries $\mathcal{C}_{ABCD}=-\mathcal{C}_{BACD}=-\mathcal{C}_{ABDC}=\mathcal{C}_{CDAB}$.
Eqs. \eqref{Weyl1} represent  the Weyl tensor $\mathcal{C}_{\text{EH}}$ of Eguchi-Hanson space and \eqref{Weyl2} are the mixed   components 
 $\mathcal{C}_{\text{mix}}$. This  Weyl tensor vanishes in the orbifold limit $a\to 0$. 

\subsection*{ Resolved   background $\M^5$ }

Similarly, we may introduce the following  Vielbein-basis for the metric \eqref{xxx}
(see  \eqref{23} and \eqref{modforms})
\ba
&e^\chi=\frac{\cos\rho}{\sqrt{1-\sin^4 a}}\dd\chi\,,\qquad e^5=\frac{1}{\sqrt{V(\rho)}} \dd \rho\,,\no \\
&e^6= \sin\rho\, \sigma_x\,,\qquad e^7= \sin\rho \,\sigma_y\,,\qquad e^8 = \sin\rho \sqrt{V(\rho)}\,\tilde\sigma_z\,,\\
&\dd s_{10}^2=\dd s^2_{AdS_5}+ e^\chi e^\chi+ \sum_{a=5}^8e^a e^a\,.
\ea
The 5-form  components  on $\mathcal{M}^5$ are then 
 $F_{\chi5678}=4$.

We find  that  the corresponding 
Weyl tensor expanded  in small $\r$ 
  is given by   $\mathcal{C}_{\text{EH}}$ \eqref{Weyl1} of the 
Eguchi-Hanson space plus  higher order  corrections 
$\order{\rho^{-4}}$. The remaining mixed components representing  $\mathcal{C}_{\text{mix}}$ are
\begin{equation}\la{D8}
\mathcal{C}_{\chi 5 \chi 5}=-\mathcal{C}_{\chi 6\chi 6}=-\mathcal{C}_{\chi 7\chi 7}=\mathcal{C}_{\chi 8\chi 8}=\frac{\sin^4 a}{\sin^4\rho}\,.
\end{equation}

\bibliographystyle{JHEP}
\baselineskip 11pt

\bibliography{N=2.bib}

\end{document}